\pdfoutput=1
\newcommand*{\ATLASLATEXPATH}{}
\documentclass[cernpreprint, english, texlive=2017]{\ATLASLATEXPATH atlasdoc}
 
\usepackage[subfigure]{\ATLASLATEXPATH atlaspackage}
\usepackage{\ATLASLATEXPATH atlaspackage}
\usepackage{\ATLASLATEXPATH atlasbiblatex}
 
\usepackage[BSM]{\ATLASLATEXPATH atlasphysics}
 
\graphicspath{{logos/}{figures/}}
 
\usepackage{multib_ew_paper-defs}

\usepackage{siunitx}

\AtlasTitle{Search for pair production of higgsinos in final states with at least three $b$-tagged jets
in $\sqrt{s} = 13$~TeV $pp$ collisions using the ATLAS~detector
}

\AtlasAbstract{
A search for pair production of the supersymmetric partners of the Higgs boson (higgsinos~$\tilde{H}$) in gauge-mediated scenarios is reported. Each higgsino is assumed to decay to a Higgs boson and a gravitino. Two complementary analyses, targeting high- and low-mass signals, are performed to maximize sensitivity.
The two analyses utilize LHC $pp$ collision data at a center-of-mass energy
$\sqrt{s} = 13$~TeV, the former with an integrated luminosity of 36.1~fb$^{-1}$ and the latter with 24.3~fb$^{-1}$, collected with the ATLAS
detector in 2015 and 2016. The search is performed in events containing missing
transverse momentum and several energetic jets, at least three of which must be identified as $b$-quark jets. No significant excess is found above the predicted background.
Limits on the cross-section are set as a function of the mass of the $\tilde{H}$ in simplified models assuming production via mass-degenerate higgsinos decaying to a Higgs boson and a gravitino. Higgsinos
with masses between 130 and 230~GeV and between 290 and 880~GeV are excluded at the 95\% confidence level.
Interpretations of the
limits in terms of the branching ratio of the higgsino to a $Z$ boson or a Higgs boson are also presented, and a 45\% branching ratio to a Higgs boson is excluded for $m_{\tilde{H}} \approx 400$ GeV.
}

\author{The ATLAS Collaboration}
 
\AtlasRefCode{SUSY-2017-02}
 
\PreprintIdNumber{CERN-EP-2018-050}
 
\AtlasJournalRef{Phys. Rev. D 98 (2018) 092002}
\AtlasDOI{10.1103/PhysRevD.98.092002}

\AtlasJournal{PRD}

\AtlasCoverAnalysisTeam{John Alison, Patrick Bryant, Trisha Farooque, Kristian Gregersen, Aurelio Juste, Michael Kagan, Nikos Konstantinidis, David Miller, Michael Nelson, Chiara Rizzi, Mel Shochet, Giordon Stark, Maximilian Swiatlowski, Loic Valery, David Wardrope}
 
\AtlasCoverEdBoardMember{Benjamin Hooberman (Chair), Otilia Ducu, Marco Battaglia}

\AtlasCoverEgroupEditors{atlas-SUSY-2017-02-editors@cern.ch}
 
\AtlasCoverEgroupEdBoard{atlas-SUSY-2017-02-editorial-board@cern.ch}
 
\hypersetup{pdftitle={ATLAS document},pdfauthor={The ATLAS Collaboration}}
 
\begin{document}
 
\maketitle

\section{Introduction}
\label{sec:intro}
 
Supersymmetry (SUSY)~\cite{Golfand:1971iw,Volkov:1973ix,Wess:1974tw,Wess:1974jb,Ferrara:1974pu,Salam:1974ig} predicts new partners of the Standard Model (SM) particles; every boson is paired with a fermionic supersymmetric partner, and vice versa. If $R$-parity conservation~\cite{Farrar:1978xj} is assumed, SUSY particles are produced in pairs and the lightest supersymmetric particle (LSP) is stable. If manifested in reality, SUSY would be a broken symmetry since the masses of the partner particles are not equal to those of the SM particles. The problem of the fine-tuning of the Higgs boson mass in the SM at the electroweak scale can be explained by the divergent diagrams canceling out their supersymmetric counterparts~\cite{Sakai:1981gr,Dimopoulos:1981yj,Ibanez:1981yh,Dimopoulos:1981zb}. These ``natural'' SUSY models generally require light partners of the gluon (gluino), top quark (stop), and the Higgs boson itself (higgsinos, $\tilde{H}_{1}^0$, $\tilde{H}^{\pm}$ and $\tilde{H}_{2}^0$)~\cite{Barbieri:1987fn}. Searches by the ATLAS and CMS collaborations have set strong limits on the masses of gluinos and stops in these models, raising the prospect that the higgsino may be light enough to be the first SUSY particle to be detected.
 
This paper presents a search for the pair production of higgsinos in models of general gauge mediation (GGM)~\cite{Meade:2008wd,Cheung:2007es,Dine:1981gu,AlvarezGaume:1981wy,Nappi:1982hm} or gauge-mediated symmetry breaking (GMSB)~\cite{Dimopoulos:1996vz,Matchev:1999ft} with a gravitino ($\gravino$) LSP, where each higgsino decays to a Higgs boson and a gravitino, in the 4-$b$-jet + \met final state.\footnote{\met is the magnitude of the missing transverse momentum vector,
which is the negative vectorial sum of the transverse momenta (\pt) of all visible particles in the event. A $b$-jet is a jet containing a hadron with a bottom quark.}
SUSY predicts five different Higgs bosons; the observed Higgs boson at $m_h \approx 125$ GeV is assumed to be the light CP-even Higgs boson ($h$) of the Minimal Supersymmetric Standard Model~\cite{Carena:2013ytb}. The high branching fraction of the observed Higgs boson to a pair of $b$-jets makes this channel particularly sensitive to these models. The search is conducted using two complementary analyses targeting high- and low-mass higgsinos. The analysis targeting the high-mass signals uses 36.1~fb$^{-1}$ of $\sqrt{s} = 13$~TeV $pp$ collision data from the LHC recorded by the ATLAS detector~\cite{PERF-2007-01} in 2015 and 2016 and utilizes \met triggers that are efficient for high-mass higgsinos. For low-mass higgsinos the \met is significantly reduced; to recover acceptance, a dedicated low-mass search inspired by the ATLAS di-Higgs resonance search \cite{ATLAS_exot_hh4b} uses a combination of $b$-jet triggers in 24.3 fb$^{-1}$ of data collected by the ATLAS detector in 2016.  This is the first search performed by ATLAS for these signatures; CMS reported similar searches at 8~\TeV~\cite{CMS-SUS-14-002} and at 13~\TeV~\cite{CMS_higgsino4b}.
 
The paper is organized as follows. The SUSY models under scrutiny are described in Section~\ref{sec:susy_sig}, followed by a brief description of the ATLAS detector in Section~\ref{sec:detector}. The datasets and simulated event samples are described in Section~\ref{sec:samples}, and the object reconstruction is summarized in Section~\ref{sec:object}. The event selection and background estimation strategies are presented for the high-mass and low-mass analyses in Sections~\ref{sec:selection_etmiss} and \ref{sec:selection_btrig}, respectively. The systematic uncertainties for both analyses are described in Section~\ref{sec:syst}, and the results are shown in Section~\ref{sec:results}. Finally, the results are interpreted in the context of model-independent upper limits on cross-sections and limits on simplified models of higgsino pair production in Section~\ref{sec:interpretation}, followed by a brief conclusion in Section~\ref{sec:conclusion}.
 
\section{SUSY signal models}
\label{sec:susy_sig}
 
In most models of supersymmetry, the higgsinos mix with gauginos (supersymmetric partners of the electroweak gauge bosons) to form mass eigenstates referred to as charginos ($\tilde{\chi}^\pm$) and neutralinos ($\tilde{\chi}^0$). Natural models often demand that the lightest neutralinos and charginos are dominated by the higgsino component. In this scenario, the masses of the four lightest such particles would be nearly degenerate~\cite{Papucci:2011wy,Barbieri:2009ev,Han:2014kaa}, with mass ordering $m_{\tilde\chi^0_1}~<~m_{\tilde\chi^\pm_1}~<~m_{\tilde\chi^0_2}$. In these models, sparticle production is dominated by the \ninoone\ninotwo, \ninoone\chinoonepm, \ninotwo\chinoonepm, and \chinoonep\chinoonem processes. In these scenarios, the heavier chargino and neutralinos can decay to the lightest neutralino (\ninoone) via off-shell $W$ and $Z$ bosons, which are assumed to decay to immeasurably low momentum particles.
 
In SUSY models with low SUSY breaking scales, such as GGM or GMSB, a nearly massless $\gravino$ is typically assumed to be the LSP; in natural models with light higgsinos, the \ninoone then becomes the next-to-lightest supersymmetric particle (NLSP). While a variety of decay scenarios is possible between the various higgsino states and the LSP, the models under study in this analysis assume that the heavier higgsinos decay first to the \ninoone and then promptly to the LSP. Depending on the specific parameters of the model, the \ninoone can decay to the \gravino via a photon, $Z$ boson, or Higgs boson~\cite{Meade:2009qv}. If \mhino\ is greater than the Higgs mass, the \ninoone is dominated by the higgsino component, and $\tan \beta$ (the ratio of expectation values of the Higgs doublets) is small, then the dominant decay would typically be via Higgs bosons, which can in turn decay to pairs of $b$-quarks, which this search targets.

\begin{figure}[htbp]
\centering
\includegraphics[width=0.3\textwidth]{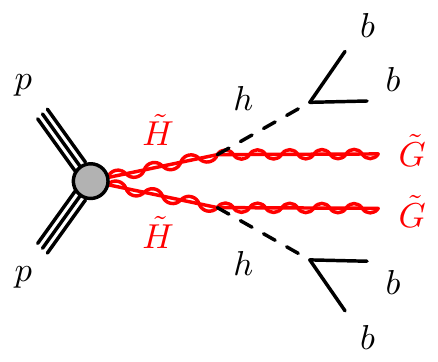}
\caption{Diagram for the simplified model considered in the analysis. The production of the \hino\ occurs
via mass-degenerate pairs of charginos or neutralinos, which decay to the \ninoone\ and immeasurably low momentum particles.}
\label{fig:feyn}
\end{figure}

These scenarios are implemented as simplified models~\cite{Alwall:2008ve,Alwall:2008ag,Alves:2011wf} as shown in Figure~\ref{fig:feyn}. The primary free parameter of the model is the mass of the degenerate higgsino states, \mhino; the mass of the LSP is set to a negligibly small value. The total signal cross-section is the sum of the four mass-degenerate higgsino pair production cross-sections.

\section{ATLAS detector}
\label{sec:detector}
 
 
The ATLAS detector is a multipurpose particle detector with a forward-backward
symmetric cylindrical geometry and nearly 4$\pi$ coverage in solid angle.\footnote{ATLAS
uses a right-handed coordinate system with its origin at the nominal interaction point in
the center of the detector. The positive $x$-axis is defined by the direction from the
interaction point to the center of the LHC ring, with the positive $y$-axis pointing
upwards, while the beam direction defines the $z$-axis. Cylindrical coordinates $(r,\phi)$
are used in the transverse plane with $\phi$ being the azimuthal angle around the $z$-axis.
The pseudorapidity $\eta$ is defined in terms of the polar angle $\theta$ by $\eta=-\ln\tan(\theta/2)$.
Rapidity is defined as $y = 0.5 \ln  [(E + p_z )/(E - p_z ) ]$  where $E$ denotes the energy and $p_z$ is the component of the momentum along the beam direction.}
The inner tracking detector (ID) consists of silicon pixel and microstrip detectors
covering the pseudorapidity region $|\eta|<2.5$, surrounded by a transition radiation tracker,
which enhances electron identification in the region $|\eta|<2.0$.
Before Run 2, a new innermost pixel layer, the insertable B-layer \cite{ATLAS-TDR-19}, was inserted at a mean sensor radius of 3.3~cm.
The ID is surrounded by a thin superconducting solenoid providing an axial 2 T magnetic field and by
a fine-granularity lead/liquid-argon (LAr) electromagnetic calorimeter covering $|\eta|<3.2$.
A steel/scintillator-tile calorimeter provides coverage for hadronic showers in
the central pseudorapidity range ($|\eta|<1.7$).
The endcaps ($1.5<|\eta|<3.2$) of the hadronic calorimeter have LAr active layers with either copper or tungsten as the absorber material.
The forward region ($3.1 < |\eta| < 4.9$) is instrumented with a LAr calorimeter for both the EM and hadronic measurements.
A muon spectrometer with an air-core toroidal magnet system surrounds the calorimeters.
Three layers of high-precision tracking chambers
provide coverage in the range $|\eta|<2.7$, while dedicated fast chambers allow triggering in the region $|\eta|<2.4$.
The ATLAS trigger system \cite{TRIG-2016-01} consists of a hardware-based level-1 trigger followed by a software-based high-level trigger.
 
\section{Data and simulated event samples}
\label{sec:samples}
 
 
The data used in this analysis were collected by the ATLAS detector from $pp$ collisions produced by the LHC at a center-of-mass-energy of 13 \tev\ with a 25 ns proton-bunch spacing during 2015 and 2016. The high-mass analyses uses data from 2015 with an integrated luminosity of 3.2 \ifb\ and from 2016 with an integrated luminosity of 32.9 \ifb, after the application of beam, detector and data-quality requirements. The low-mass analysis uses data from 2016 with an integrated luminosity of 24.3 \ifb. The uncertainties in the integrated luminosities are $\pm2.1$\% and $\pm2.2$\% for the 2015 and 2016 datasets, respectively, determined from a calibration of the luminosity scale using $x$--$y$ beam-separation scans performed in August 2015 and May 2016, following a methodology similar to that detailed in Ref.~\cite{DAPR-2013-01}. The difference in luminosity between the analyses is due to using different triggers. In the high-mass analysis, events are required to pass an $\met$ trigger with thresholds of 70~\gev, 90~\gev, 100~\gev, and 110~\gev\ in the high-level trigger for the 2015, and early, mid, and late 2016 datasets, respectively. These triggers are fully efficient for events passing the preselection defined in Section~\ref{sec:selection_etmiss}, which requires the offline reconstructed $\met$ to exceed 200 GeV. In the low-mass analysis, a combination of three triggers requiring $b$-tagged jets are used. These require events to feature either one $b$-tagged jet with $\pt>225$ \gev, two $b$-tagged jets with $\pt>55$ \gev\ and one additional jet with $\pt>100$ \gev, or two $b$-tagged jets with $\pt>35$ \gev\ and two additional jets with $\pt>35$ \gev. During the 2016 data taking, a fraction of the data suffered from faulty vertex reconstruction, and those events were not retained.  For the combined 2015 and 2016 dataset, there are an average of 24 inelastic $pp$ collisions per bunch crossing (interactions other than the hard scatter are referred to as ``pileup'').

Samples of Monte Carlo (MC) simulated events are used to model the signal and background processes in the high-mass analysis, except multijet processes, which are estimated from data. In the low-mass analysis, the background is dominated by multijet processes that are not modeled reliably in simulation, and the estimation methodology is thus based on data control samples as described in Section~\ref{sec:selection_btrig}.  The SUSY signal
samples 
were generated with up to two
additional partons using \MGMCatNLO~\cite{Alwall:2014hca} v2.3.3 at leading order (LO) with the
NNPDF 2.3~\cite{Ball:2012cx} parton distribution function (PDF) set. These
samples were interfaced to \PYTHIA v8.186~\cite{Sjostrand:2007gs}
for the modeling of the parton showering, hadronization and underlying event.
 
The generators used to simulate signal processes for both analyses and background processes for the high-mass analysis are described in Table~\ref{t-MCsamples}. The dominant background is \ttbar production, which was simulated with the \POWHEGBOX~\cite{Alioli:2010xd}~v2 event generator. The $Wt$- and s-channel production of single top quarks was also simulated with this generator, but t-channel prodution was simulated with \POWHEGBOX~v1. Backgrounds from $W/Z$+jets processes were simulated using the \SHERPA v2.2.0~\cite{Gleisberg:2008ta} event
generator, while \SHERPA v2.1.1 was used to simulate diboson
production processes. The production of \ttbar\ pairs in association with a $W$, $Z$, or Higgs boson was modeled by samples generated using \MGMCatNLO~\cite{ATL-PHYS-PUB-2016-005}. \MGMCatNLO was also used to simulate $\ttbar \ttbar$ production. All details of the versions of the generators, showering models, sets of tuned parameters, and PDF sets are given in Table~\ref{t-MCsamples}.
 
All background processes are initially normalized using the best available theoretical calculation for their
respective cross-sections; the \ttbar\ contribution is further normalized using data as described in Section~\ref{sec:selection_etmiss}. The order of this calculation in perturbative QCD for each process is listed in
Table~\ref{t-MCsamples}.

The signal samples are normalized using the best cross-section calculations at NLO in the strong coupling
constant, adding the resummation of soft gluon emission at next-to-leading-logarithm (NLL) accuracy~\cite{Fuks:2012qx,Fuks:2013vua}. The generator, set of tuned parameters, and PDF set are described in Table~\ref{t-MCsamples}. The nominal cross-section
and its uncertainty are taken from an envelope of cross-section predictions using different PDF sets and factorization
and renormalization scales, as described in Ref.~\cite{Borschensky:2014cia}. The cross-section of higgsino pair production at $\mhino = 150$ GeV is 3830 $\pm$ 160 fb, while at $\mhino = 900$ GeV it is 1.8 $\pm$ 0.2 fb.
 
All simulated event samples were processed with the full ATLAS detector simulation~\cite{SOFT-2010-01} using \MYGEANT4~\cite{Agostinelli:2002hh},
with the exception of signal samples,
which were processed with a fast simulation~\cite{ATL-PHYS-PUB-2010-013} that uses a
parameterization of the calorimeter response and \MYGEANT4 for the
ID and the muon spectrometer response.  Pileup collisions were simulated with \PYTHIA 8~\cite{Sjostrand:2007gs} and overlaid on each MC event. Weights are assigned to the simulated events such that the distribution of the number of pileup interactions in the simulation matches the corresponding distribution in the data.
The simulated events were reconstructed with the same algorithms used for data.

\begin{table}[t]
 
\centering
{\footnotesize
\caption{Event generators used for the different
processes. Information is given about the underlying-event
sets of tuned parameters, the PDF sets, and the pQCD highest-order accuracy
used for the normalization of the different samples.}
\label{t-MCsamples}
\begin{tabular}{ccccc}
\toprule
Process &    Event Generator                   & Tune set & PDF set  &  Cross-section  \\
& + fragmentation/hadronization  &      &          &  order \\
\midrule
\textbf{SUSY signal} &  \MGMCatNLO v2.3.3
& {\textsc A14} & {\textsc NNPDF2.3} & NLO+NLL~\cite{Beenakker:1996ch,Kulesza:2008jb,Kulesza:2009kq,Beenakker:2009ha,Beenakker:2011fu,Borschensky:2014cia} \\
& + \PYTHIA v8.186 & & & \\
\hline
\textbf{\ttbar} & \POWHEGBOX v2
& {\textsc PERUGIA2012} & {\textsc CT10} & NNLO+NNLL~\cite{Czakon:2011xx} \\
& + \PYTHIA v6.428 & & & \\
\hline
\textbf{Single top} & \POWHEGBOX v1 or v2
& {\textsc PERUGIA2012} & {\textsc CT10} & NNLO+NNLL~\cite{Kidonakis:2011wy,Kidonakis:2010ux,Kidonakis:2010tc}  \\
& + \PYTHIA v6.428 & & & \\
\hline
\textbf{${\ttbar}W$/${\ttbar}Z$/$\ttbar \ttbar$} & \MGMCatNLO v2.2.2
& {\textsc A14} & {\textsc NNPDF2.3} & NLO~\cite{deFlorian:2016spz} \\
& + \PYTHIA v8.186 & & & \\
\hline
\textbf{${\ttbar}H$} & \MGMCatNLO v2.2.1
& {\textsc UEEE5} & {\textsc CT10} & NLO~\cite{Heinemeyer:2013tqa} \\
& + \MYHERWIG++ v2.7.1 & & & \\
\hline
\textbf{Diboson}  & \SHERPA v2.1.1 & Default & {\textsc CT10} & NLO~\cite{ATL-PHYS-PUB-2016-002} \\
$WW$, $WZ$, $ZZ$ & & & & \\
\hline
\textbf{$W$/${Z}+$jets} & \SHERPA v2.2.0 & Default & {\textsc NNPDF3.0} & NNLO~\cite{Catani:2009sm}  \\
\bottomrule
\end{tabular}
}
\end{table}

 
\section{Object reconstruction}
\label{sec:object}
 
 
Interaction vertices are reconstructed from at least two tracks with $\pt > 0.4~\gev$, and are required to be consistent with the beamspot envelope. The primary vertex is identified as the one with the largest sum of squares of the transverse momenta from associated tracks ($\sum{p_{\mathrm{T,track}}^{2}}$)~\cite{ATL-PHYS-PUB-2015-026}.

Jets are reconstructed from three-dimensional topological energy clusters~\cite{PERF-2014-07} 
in the calorimeter using the anti-$k_{t}$ jet algorithm~\cite{Cacciari:2008gp, Cacciari:2011ma} with a radius parameter of 0.4. Each topological cluster is calibrated to the electromagnetic scale response prior to jet reconstruction. 
The reconstructed jets are then calibrated to the particle level by the application of a jet energy scale derived from $\sqrt{s}= 13~\tev$ data and simulations~\cite{Aaboud:2017jcu}. Quality criteria are imposed to reject events that contain at least one jet arising from non-collision sources or detector noise~\cite{ATLAS-CONF-2015-029}. To reject jets with $|\eta| < 2.4$ that originate from pileup interactions, further requirements are applied by means of a multivariate algorithm using information about the tracks matched to each jet~\cite{PERF-2014-03}. Candidate jets are required to have \pt$>$~20 \gev\ and $|\eta| < 2.8$ in the high-mass analysis and \pt$>$~25 \gev\ and $|\eta| < 2.5$ in the low-mass analysis. After resolving overlaps with electrons and muons, as described below, selected jets are required to satisfy the stricter requirement of \pt$>$~25~\gev\ in the high-mass analysis and 40~\gev\ in the low-mass analysis. The higher \pt requirement in the low-mass analysis is the result of the $b$-jet trigger thresholds.
 
A candidate jet is tagged as a $b$-jet by a multivariate algorithm using information about the impact parameters of ID tracks matched to the jet, the presence of displaced secondary vertices, and the reconstructed flight paths of $b$- and $c$-hadrons inside the jet~\cite{PERF-2012-04,ATL-PHYS-PUB-2016-012}. The $b$-tagging working point with an efficiency of 77\% to identify $b$-jets with \pt$>$~20 \gev, as  determined from a sample of simulated \ttbar events, is optimal in the high-mass analysis, while the low-mass analysis uses a tighter working point with 70\% $b$-tagging efficiency to suppress the large contribution from light-flavor jets in the multijet background. The corresponding rejection factors against jets originating from $c$-quarks, $\tau$-leptons and light quarks and gluons in the same sample for the selected working point are 6, 22, and 134, respectively, for the high-mass analysis and 12, 55, and 381, respectively, for the low-mass analysis.
 
Electron candidates are reconstructed from energy clusters in the electromagnetic calorimeter and ID tracks and are required to have $|\eta|<$~2.47 and satisfy a set of ``loose'' quality criteria~\cite{ATLAS-CONF-2016-024, PERF-2013-05}. Muon candidates are reconstructed from matching tracks in the ID and muon spectrometer. They are required to meet ``medium'' quality criteria, as described in Ref.~\cite{PERF-2015-10}, and to have $|\eta|<$~2.5.
An isolation requirement is applied to both the electrons and muons, and is based on the scalar sum of \pt of additional ID tracks in a cone around the lepton track. This isolation requirement is defined to ensure a constant efficiency of around 99\% across the whole electron transverse energy and muon transverse momentum ranges measured in $Z\rightarrow \ell^+\ell^-$ events \cite{ATLAS-CONF-2016-024,PERF-2013-05,PERF-2015-10}. The average angular separation between the lepton and the $b$-jet in semileptonic top quark decays narrows as the \pt of the top quark increases. This increased collimation is accounted for by setting the radius of the isolation cone to $\Delta R = \mathrm{min}(0.2, 10~\gev/p_{\mathrm{T}}^{\mathrm{lep}}$), where $p_{\mathrm{T}}^{\mathrm{lep}}$ is the lepton \pt.
In the high-mass analysis, the selected electrons are further required to meet the ``tight'' quality criteria~\cite{ATLAS-CONF-2016-024, PERF-2013-05}. 
Leptons are used in the calculation of \met, in the four-momentum correction of $b$-tagged jets, and to resolve overlaps between each other and with jets. These leptons are required to have \pt~$>$~5~\gev, while for vetoing events, the leptons are required to have \pt~$>$~20~\gev.
 
Overlaps between candidate objects are removed sequentially. If a reconstructed muon shares an  ID track with an
electron, the electron is removed. In the high-mass analysis, any non-$b$-tagged jet whose axis lies within ${\Delta}R = 0.2$ of an electron is removed.\footnote{${\Delta}R = \sqrt{(\Delta y)^{2} + (\Delta\phi)^{2}}$  defines the distance in rapidity $y$ and azimuthal angle $\phi$.} Any electrons reconstructed within ${\Delta}R = \mathrm{min}(0.4, 0.04 + 10~\gev/\pt)$ of the axis of any surviving jet are removed. If a non-$b$-tagged jet is reconstructed within ${\Delta}R = 0.2$ of a muon and the jet has fewer than three associated tracks or the muon energy constitutes most of the jet energy, then the jet is removed.  Muons reconstructed within a cone of size ${\Delta}R = \mathrm{min}(0.4, 0.04 + 10~\gev/\pt)$ around the axis of any surviving jet are removed. The same overlap procedure is applied in the low-mass analysis for jets, muons and electrons, except that $b$-tagged jets are treated the same way as non-$b$-tagged jets.
 
To account for the presence of $b$- and $c$-hadron decays to muons, which do not deposit their full energy in the calorimeter, a correction is applied to $b$-tagged jets if a muon is found within $\Delta R = 0.4$ of the jet axis before the overlap removal. The correction consists in adding the  muon four-momentum to that of the jet, and removing the energy deposited by the muon in the calorimeter. If more than one muon is found, the one closest to the jet axis is used.
 
The missing transverse momentum \met\ in the event is defined as the magnitude of the negative vector sum ($\vec{\pt}^{\mathrm{miss}}$) of the transverse momenta of all selected and calibrated electrons, muons, and jets in the event with an extra term added to account for energy deposits that are not associated with any of these objects. This ``soft'' term is calculated from ID tracks matched to the primary vertex (and not matched to any of the objects building \met), making it more resilient to contamination from pileup interactions ~\cite{ATL-PHYS-PUB-2015-023, ATL-PHYS-PUB-2015-027}.
 
Corrections derived from data control samples are applied to simulated events to account for differences between data and simulation in the reconstruction efficiencies, momentum scale, and resolution of leptons; in the $b$-tagging efficiency for $b$-jets and mistag rates for non-$b$-jets; and in the efficiency for rejecting jets originating from pileup interactions. In the low-mass analysis, corrections are applied to account for mismodeling of the $b$-jet trigger efficiencies in the simulation.
 
 
\section{Event selection and background estimation}
For the high-mass analysis, events are selected using \met triggers. Events with at least three $b$-jets are further analyzed, and jet pairs are assigned to two Higgs candidates. The dominant \ttbar\ background is suppressed by requirements on the kinematic variables related to the visible and invisible energy in the event. Several exclusive signal regions (SR) are defined to target a wide range of higgsino masses. Control regions (CR) and validation regions (VR) are defined for each SR by inverting requirements on the reconstructed Higgs boson mass and relaxing kinematic requirements. The backgrounds are estimated from MC simulation, after normalizing to data in the CRs and ensuring reliable background modeling in the VRs.
 
For the low-mass analysis, events are selected with a combination of $b$-jet triggers, and events with four $b$-jets are further analyzed by grouping the jets into Higgs candidates. A purely data-driven background estimate uses sidebands in the Higgs boson mass to estimate the background in the signal region, while further validation regions in the sidebands validate the background modeling. The search is performed by constructing exclusive signal regions binned in the visible and invisible energy in the event.
 
Two classes of signal regions are defined for each of the two analyses. Discovery regions are optimized to maximize the expected discovery power for benchmark signal models and to facilitate the reinterpretation of results. These SRs are defined to probe the existence of a signal or to assess model-independent upper limits on the number of signal events. To maximize exclusion sensitivity to a variety of signal models, a further set of fully orthogonal signal regions is also constructed; the result of a combined fit across all these regions is significantly stronger than that to a single region because information about the expected shape of the signal for different variables provides additional constraining power.
 
\subsection{High-mass analysis}
\label{sec:selection_etmiss}
 
\subsubsection{Event selection}
\label{high_event_selection}
One of the key elements of the analysis is the identification of the Higgs bosons originating from the higgsino decays. To choose which jets are used in the reconstruction of the Higgs boson candidates, the following ordered criteria are used. If there are exactly four $b$-tagged jets in the analysis, those four are used. If there are more than four $b$-tagged jets, the four with the highest \pt are used. If there are three $b$-tagged jets and at least one untagged jet, the three tagged jets and the untagged jet with the highest \pt are used.
 
To determine the optimal pairing of jets, the quantity $\dRmax = \mathrm{max}(\Delta R(h_1), \Delta R(h_2))$ is minimized, where $\Delta R(h)$ is the distance in $\eta-\phi$ space between the jets constituting a Higgs boson candidate.
This selection efficiently reconstructs decays of both the Higgs and $Z$ bosons to $b$-jets, giving sensitivity to final states where the branching ratio of higgsino decays to Higgs bosons is not 100\%.

The following variables, constructed from the selected jets and the \ptmiss of the event, are used to discriminate between the signal and various backgrounds. The effective mass is defined as the scalar sum of the \pt of the four jets used in the Higgs boson reconstruction and the $\met$: $\meffb = \sum_{i=1,..,4} {\pt}^{j_i} + \met$. The minimum $\Delta \phi$ between any of the leading four jets and \ptmiss, $\dphimin = \textrm{min}(|\phi_1 - \phi_{\ptmiss}|,...,|\phi_4 - \phi_{\ptmiss}|)$, suppresses multijet backgrounds arising from mismeasured jets. The minimum transverse mass between the \ptmiss\ and the three leading \bjets,
$\mtb = \allowbreak \mathrm{min}_{i\leq 3}  \sqrt{(\met+p_\textrm{T}^{j_i})^2 - (p_{x}^{\mathrm{miss}}+p_x^{j_i})^2 - (p_{y}^{\mathrm{miss}}+p_y^{j_i})^2 }$, has a kinematic endpoint near the top mass for $t\bar{t}$ backgrounds, while the value of \mtb can be much larger in signal processes.  The \njet and \nbjet variables are the number of selected signal jets and $b$-jets, respectively. The masses of the higher- and lower-mass candidate Higgs bosons are $m(h_1)$ and $m(h_2)$.

Preselection criteria for the high-mass analysis
require $\met > 200$~\gev, in addition to the
$\met$ trigger requirement, and at least four jets of which at least three
must be $b$-tagged. The events are required to have no selected leptons, and $\dphimin > 0.4$. The data and the predicted background are found to agree well at the preselection level, as shown in Figure~\ref{fig:presel_highmet}. Selected signal models are overlaid for comparison.
 
\begin{figure}[htbp]
\centering
\includegraphics[width=0.650\textwidth]{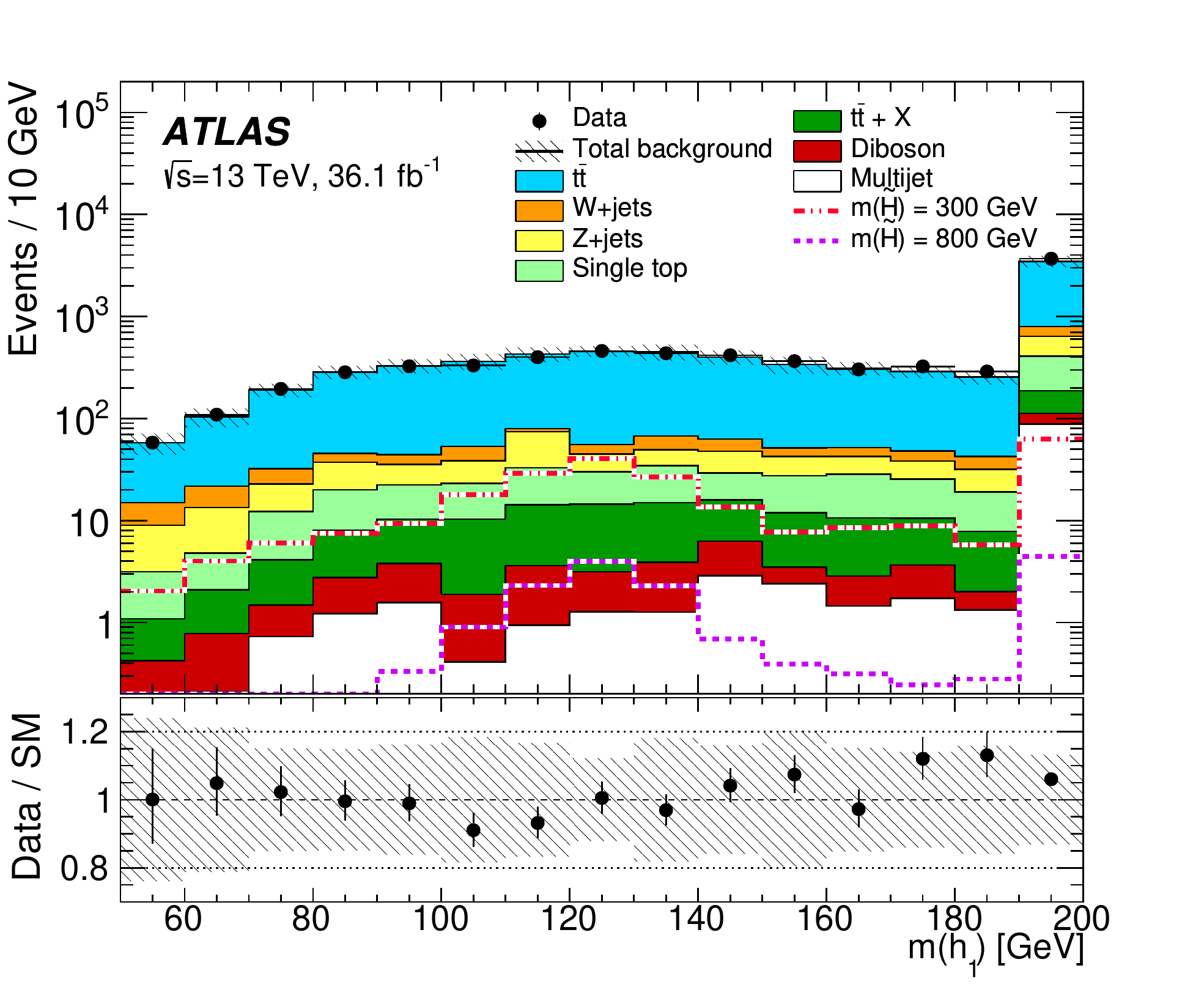}\\
\includegraphics[width=0.650\textwidth]{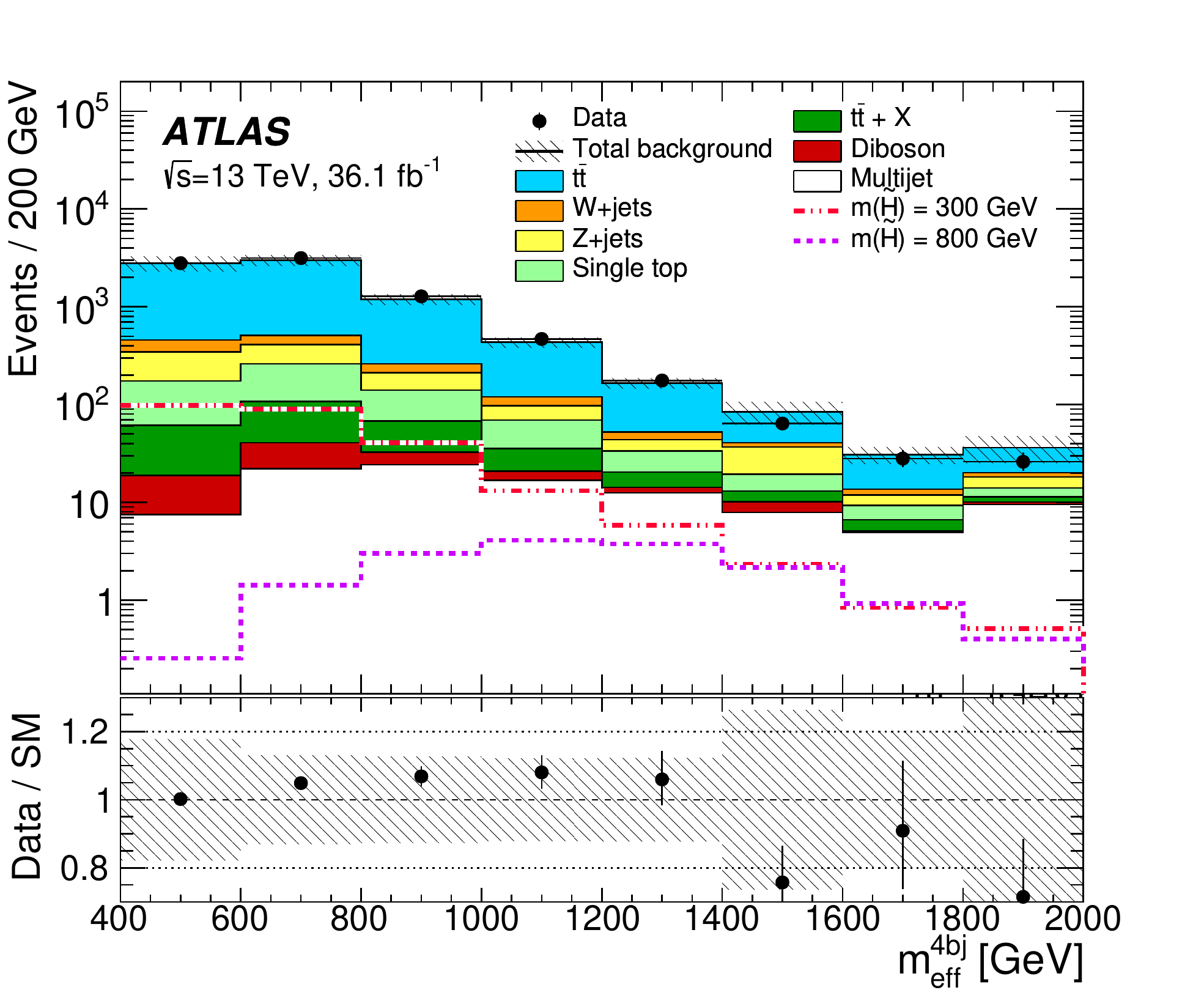}
\caption{Distributions of $m(h_1)$ (top) and \meffb (bottom) for events passing the preselection criteria of the high-mass analysis. All backgrounds (including  $\ttbar$) are
normalized using the best available theoretical calculation described in Section~\ref{sec:samples}. The dashed histograms show the distributions of the variables for selected signal models at the best available theoretical cross section. The statistical and experimental systematic uncertainties (as defined in Section~\ref{sec:syst_high}) are included in the uncertainty band. The last bin includes overflows.
}
\label{fig:presel_highmet}
\end{figure}

To enhance the sensitivity to the various signal benchmarks described in Section \ref{sec:susy_sig},
multiple signal regions (SRs) are defined.  Seven fully orthogonal signal regions optimized for exclusion sensitivity are defined in Table~\ref{tab:SR}. The regions are defined by $b$-jet multiplicity, \dRmax, and \meffb. Requirements on \mtb and \njet are optimized within each of these bins separately. All signal regions require $\met > 200$ GeV so that the trigger is efficient, and all require $\dphimin > 0.4$ to suppress backgrounds from multijet production. The names of the signal regions are defined as SR-X-meffY-Z: X can be 3b or 4b and defines the $b$-jet multiplicity; Y $\in \{1,2,3\}$ defines the particular bin in \meffb; and Z $\in\{$A,B$\}$ defines the \dRmax bin.
 
\begin{table}[h]
\caption{Signal region definitions for the high-mass analysis. The units of \met, \mtb, $m(h_1)$, $m(h_2)$, and \meffb are GeV. These variables are defined in Section~\ref{high_event_selection}.}
\begin{center}
\resizebox{1.\textwidth}{!}{
\begin{tabular}{|l|c|c|c|c|c|c|c|c|}
\hline
& SR-3b-meff1-A & SR-3b-meff2-A & SR-3b-meff3-A & SR-4b-meff1-A & SR-4b-meff1-B & SR-4b-meff2-A & SR-4b-meff2-B  & SR-4b-meff1-A-disc \\
\hline
\nbjet &  $=$3 &  $=$3 &  $\geq$3 &  $\geq$4 &  $\geq$4 &  $\geq$4 &  $\geq$4  & $\geq4$\\
\hline
\met & \multicolumn{8}{|c|}{$>$ 200}\\
\hline
\dphimin    & \multicolumn{8}{|c|}{$>$0.4}\\
\hline
\njet &  4--5 &  4--5 &  4--5 &  4--5 &  4--5 &  4--6 &  4--6 & 4--5\\
\hline
\mtb &  $>$150 &  $>$150 &  $>$130 & - & - & - & - & - \\
\hline
$m(h_1)$ &    \multicolumn{8}{|c|}{110--150}\\
\hline
$m(h_2)$ &    \multicolumn{8}{|c|}{90--140}\\
\hline
\dRmax &  0.4--1.4 &  0.4--1.4 &  0.4--1.4 &  0.4--1.4 &  1.4--2.4 &  0.4--1.4 &  1.4--2.4 & 0.4--1.4 \\
\hline
\meffb &  600--850 &  850--1100 &  $>$1100 &  600--850 &  600--850 &  850--1100 &  850--1100 & $>600$ \\
\hline
\end{tabular}
}
 
\label{tab:SR}
\end{center}
\end{table}

While the previously described regions are optimized to maximize exclusion sensitivity to particular models, the \meffb binning in some cases reduces the signal contribution in individual bins, thereby reducing the discovery sensitivity. For this reason, two single-bin SRs, targeting medium- and high-mass  higgsinos, are optimized for discovery. At intermediate mass, the most sensitive region modifies SR-4b-meff1-A by removing the upper bound on \meffb; this region is called SR-4b-meff1-A-disc and is also defined in Table~\ref{tab:SR}. At high mass, the SR-3b-meff3-A already has no upper bound on \meffb and is therefore already a region with strong discovery sensitivity. Both of these regions are defined to probe the existence of a signal and in its absence to assess model-independent upper limits on the number of signal events.
 
All aspects of the SR definitions, including the choice of Higgs boson reconstruction algorithm and the variables used in the analysis together with their associated cuts, were optimized using simulated events.
 
\subsubsection{Background estimation strategy}

The main background in the SRs is the production of a \ttbar\ pair in association with heavy- and light-flavor
jets. A normalization factor for this background is extracted  for each SR from a data control region (CR)
that has comparable background composition and kinematics,
ensured by using similar kinematic
requirements in the two regions.
The CRs and SRs are defined to be mutually exclusive by binning in $m(h_1)$ and $m(h_2)$, as shown in Figure~\ref{fig:binning_crvr}.
Signal contributions in the CRs are suppressed by choosing events with Higgs boson candidate masses far from the SM value, leading to a signal contamination in the CRs of 10\% at most. Requirements on variables such as \mtb are loosened in order to provide enough events in the CR to provide a meaningful normalization.  
The $\ttbar$ normalization is cross-checked
in validation regions (VRs) similar in background composition
to the SR.   
 
\begin{figure}[htbp]
\centering
\includegraphics[width=0.490\textwidth]{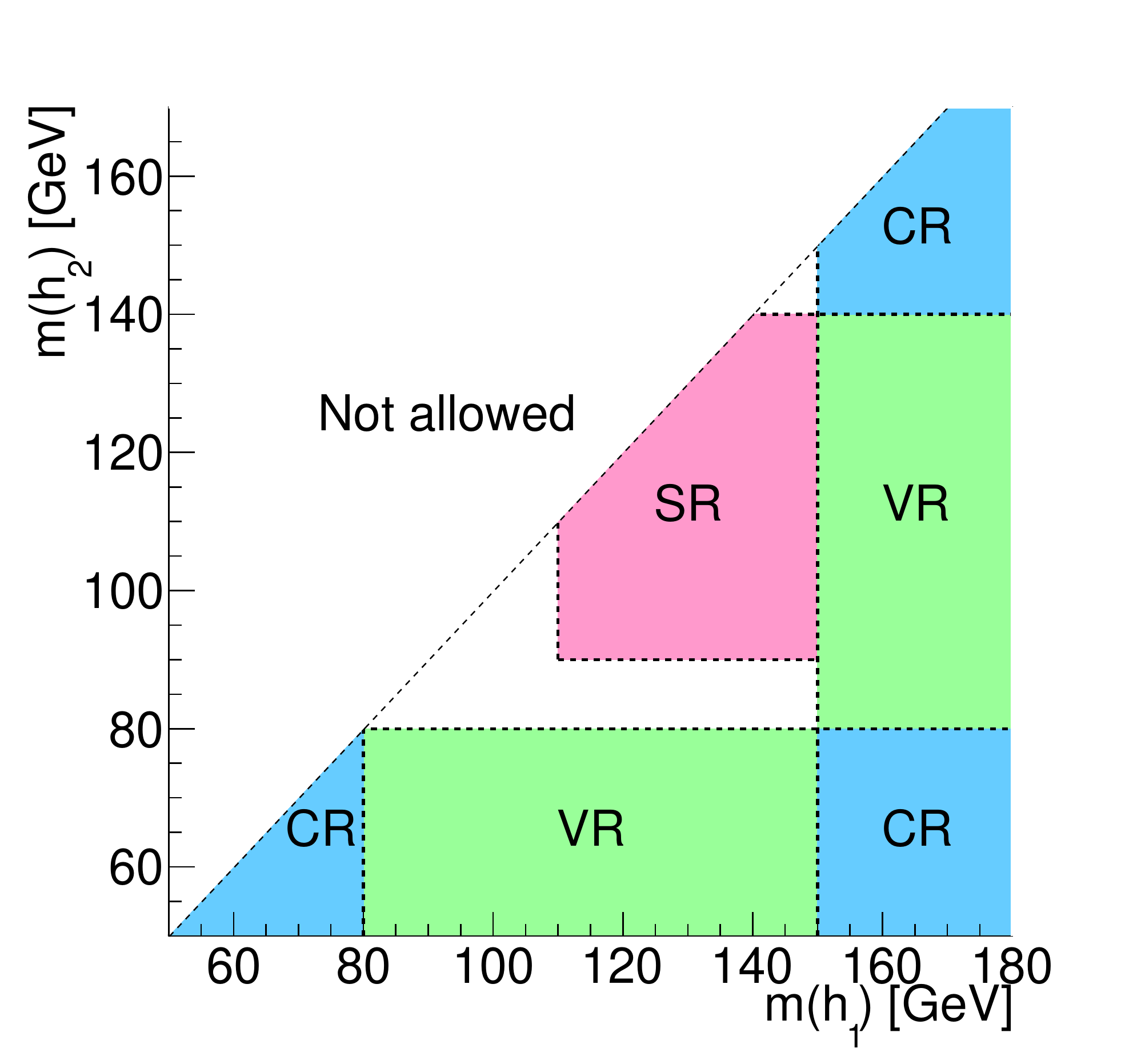}
\caption{The division of signal, control, and validation regions using the $m(h_1)$ and $m(h_2)$ variables in the high-mass analysis.}
\label{fig:binning_crvr}
\end{figure}

The non-\ttbar\ backgrounds consist mainly of single-top, $W$+jets,
$Z$+jets, \ttbar+$W/Z/h$, $t\bar{t}t\bar{t}$ and diboson events.
The shape of each distribution for these processes is taken from the
simulation, and they are normalized using the best available theory prediction.
The multijet background is very small or negligible in all regions. It
is estimated
using a procedure described in Ref. \cite{Aad:2012fqa}, in which  the jet
response is determined from
simulated dijet events and tuned to data. The response function and corrections are derived separately for $b$-tagged jets. This response function is then used to smear
the \pt of jets in multijet events from data with low \met-significance, defined as $\met / \sqrt{\sum_i p_\textrm{T}^i}$, where the sum is over all jets in the event.
The smeared predictions are normalized to the recorded luminosity using a control region with $\dphimin < 0.1$, where the \met is directly attributable to mismeasurement of one of the jets. The results are validated with data in the region $0.1 < \dphimin < 0.4$.

Control regions used to normalize the \ttbar background are constructed to be as similar to the signal regions as possible, although requirements on \mtb are relaxed to increase the statistical precision in the control region. The control regions are made orthogonal to the signal regions by changing the mass requirement on the Higgs boson candidates. Each \meffb bin of the SR has a corresponding CR; bins in \dRmax are combined to increase the statistical power of the control regions. The names of the control regions follow those of the signal regions and are summarized in Table~\ref{tab:CR}. Because the discovery region SR-4b-meff1-A-disc is nearly the same as the SR-4b-meff1-A region, CR-4b-meff1 is used to normalize both. The values of the normalization factors, the expected numbers of background events, and
the observed data yields in all the CRs of the high-mass analysis are shown in Figure~\ref{fig:pullCR}.
 
\begin{figure}[htbp]
\centering
\includegraphics[width=0.9\textwidth]{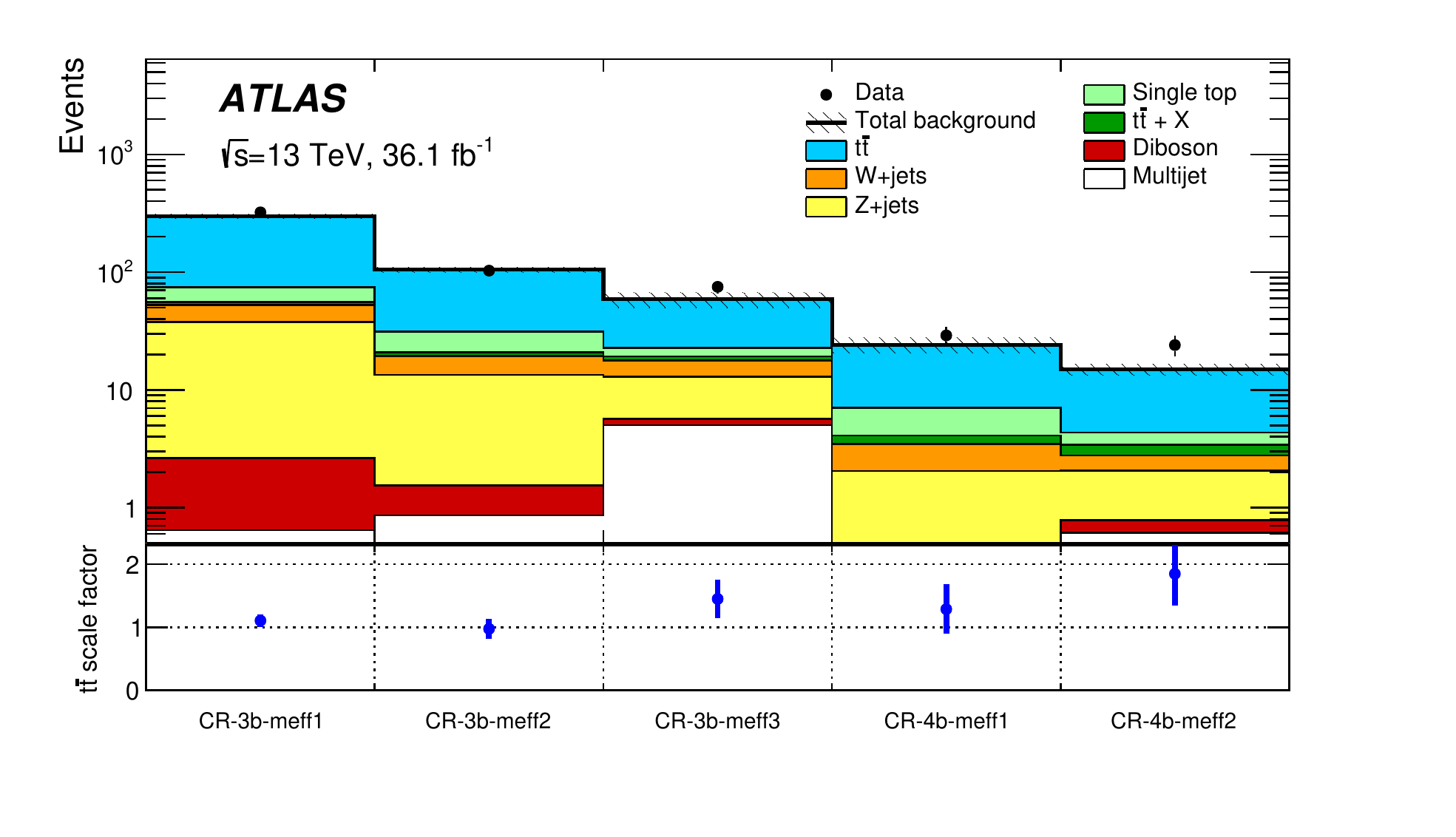}
\caption{Event yields in control regions and related \ttbar\
normalization factors after the background-only fit for
the high-mass analysis. The upper panel shows
the observed number of events and the predicted background yield before the fit.
All uncertainties described in Section \ref{sec:syst_high} are included in the uncertainty band. The background category $\ttbar+X$ includes $\ttbar W/Z$, $\ttbar H$, and $\ttbar \ttbar$ events.
The $\ttbar$ normalization is obtained from the fit
and is displayed in the bottom panel.
}
\label{fig:pullCR}
\end{figure}

Finally, the validation regions are used to measure the accuracy of the control region normalizations. They are made orthogonal to the signal and control regions by changing the mass requirement on the Higgs boson candidates, using the low-mass sideband of $m(h_2)$ and the high-mass sideband of $m(h_1)$ as shown in Figure~\ref{fig:binning_crvr}. To accept more events, the \mtb and \dRmax requirements are loosened, and the \meffb requirements are lowered in some cases as well. The full definitions are shown in Table~\ref{tab:VR}. The signal contamination in the VRs for signals near the limit of sensitivity is found to be less than 10\%.

\begin{table}[h]
\caption{Control region definitions in the high-mass analysis. The units of \met, \mtb, $m(h_1)$, $m(h_2)$, and \meffb are GeV. These variables are defined in Section~\ref{high_event_selection}.}
\begin{center}
\resizebox{0.75\textwidth}{!}{
\begin{tabular}{|l|c|c|c|c|c|}
\hline
& CR-3b-meff1 & CR-3b-meff2 & CR-3b-meff3 & CR-4b-meff1 & CR-4b-meff2 \\
\hline
\nbjet &  $=$3 &  $=$3 &  $\geq$3 &  $\geq$4 &  $\geq$4 \\
\hline
\met  & \multicolumn{5}{|c|}{$>$ 200}\\
\hline
\dphimin  & \multicolumn{5}{|c|}{$>$0.4}\\
\hline
\njet &  4--5 &  4--5 &  4--5 &  4--5 &  4--6 \\
\hline
\mtb &  $>$100 &  $>$100 &  $>$100 & - & - \\
\hline
$m(h_1)$, $m(h_2)$  &  \multicolumn{5}{|c|}{ ($m(h_1)<$80, $m(h_2)<$80) or ($m(h_1)>$150, $m(h_2)<$80) or ($m(h_1)>$150, $m(h_2)>$140)    }\\
\hline
\dRmax &  0.4--4 &  0.4--4 &  0.4--4 &  0.4--4 &  $\geq$ 0.4 \\
\hline
\meffb &  600--850 &  850--1100 &  $>$1100 &  600--850 &  850--1100 \\
\hline
\end{tabular}
}
\label{tab:CR}
\end{center}
\end{table}
 
\begin{table}[h]
\caption{Validation region definitions in the high-mass analysis. The units of \met, \mtb, $m(h_1)$, $m(h_2)$, and \meffb are GeV. These variables are defined in Section~\ref{high_event_selection}.}
\begin{center}
\resizebox{1\textwidth}{!}{
\begin{tabular}{|l|c|c|c|c|c|c|c|}
\hline
& VR-3b-meff1-A & VR-3b-meff2-A & VR-3b-meff3-A & VR-4b-meff1-A & VR-4b-meff1-B & VR-4b-meff2-A & VR-4b-meff2-B \\
\hline
\nbjet &  $=$3 &  $=$3 &  $\geq$3 &  $\geq$4 &  $\geq$4 &  $\geq$4 &  $\geq$4 \\
\hline
\met  &  \multicolumn{7}{|c|}{$>$200}\\
\hline
\dphimin &  \multicolumn{7}{|c|}{$>$0.4}\\
\hline
\njet &  4--5 &  4--5 &  4--5 &  4--5 &  4--5 &  4--6 &  4--6 \\
\hline
\mtb  & $>$120   & $>$100  & $>$80  & -  &  - & -  & -  \\
\hline
$m(h_1)$, $m(h_2)$  &  \multicolumn{7}{|c|}{   (80$<m(h_1)<$150, $m(h_2)<$80) or ($m(h_1)>$150, 90$<m(h_2)<$140)   }\\
\hline
\dRmax &  0.4--1.5 &  0.4--1.7 &  0.4--1.7 &  0.4--1.7 &  1.4--3 &  0.4--1.7 &  1.4--3 \\
\hline
\meffb   & 550--900   & 800--1150  & $>$1050  & 550--900  & 550--900  & 800--1150  & 800--1150  \\
\hline
\end{tabular}
}
\label{tab:VR}
\end{center}
\end{table}
 
The expected SM background is determined separately in each SR from
a profile likelihood fit \cite{Cowan:2010js} implemented in the
HistFitter framework \cite{HFpaper}, referred to as a background-only fit. The fit uses
as a constraint the observed event yield in the associated CR to adjust the
$\ttbar$ normalization, assuming that no signal contributes to this yield, and applies
that normalization factor to the number of $\ttbar$ events predicted by simulation in
the SR.
 
The inputs to the fit for each SR are the numbers of events
observed in its associated CR and the numbers of events predicted by
simulation in each region for all background processes. The numbers of observed and
predicted events in each CR are described by Poisson probability density
functions. The systematic uncertainties, described in Section~\ref{sec:syst_high}, in the expected values are included in the fit
as nuisance parameters. They are constrained by Gaussian distributions with widths
corresponding to the sizes of the uncertainties and are treated as correlated, when
appropriate, between the various regions. The product of the various probability density
functions forms the likelihood, which the fit maximizes by adjusting the $\ttbar$
normalization and the nuisance parameters.

 
\subsection{Low-mass analysis}
\label{sec:selection_btrig}
\subsubsection{Event selection}\label{sec:selection_lowmass}
 
The low-mass analysis targets events with reduced \met where the high-mass analysis has no sensitivity. Events are required to have at least four $b$-tagged jets. If more than four jets in the event are $b$-tagged, the four jets with the highest $b$-tagging score are used. When forming the Higgs candidates from the four jets, a weak requirement on the maximum $\Delta R$ separation of the jets is imposed as a function of the invariant mass of the di-Higgs system.
After applying this selection, the optimal pairing of the jets into Higgs candidates is achieved by minimizing the quantity $D_{hh}$, defined as:
\begin{eqnarray*}
D_{hh} &=& \left|m_{2j}^\textrm{lead} - \frac{120}{110}m_{2j}^\textrm{subl}\right|,
\end{eqnarray*}
where $m_{2j}^\textrm{lead}$ and $m_{2j}^\textrm{subl}$ are the masses of the leading and subleading Higgs boson candidates, respectively.
This definition is consistent with pairing the jets into two Higgs boson candidates of roughly equal mass. The subleading mass is scaled by the ratio of the median values of the narrowest intervals in $m_{2j}^\textrm{lead}$ and $m_{2j}^\textrm{subl}$ that contain 90\% of the signal in simulations. 
The pairing used in the high-mass analysis, which combines the $b$-tagged jets with the smallest $\Delta R$ separation into Higgs boson candidates, is sub-optimal for the low-mass analysis. This is because the Higgs bosons from low-mass signals have lower \pt, resulting in a larger $\Delta R$ separation of the $b$-quarks from their decays. 
 
After selecting the two Higgs boson candidates, the background mostly consists of multijet events and a small contribution from \ttbar\ production. The \ttbar\ background consists of hadronic \ttbar\ events at low \met and leptonic \ttbar\ events at high \met. For approximately 50\% of the leptonic \ttbar\ events, the two Higgs boson candidates are formed from the two $b$-jets from the top quark decays and a $b\bar{b}$ pair from initial-state radiation. For the hadronic \ttbar\ and the remaining leptonic \ttbar\ events, the Higgs candidates are predominantly formed from different combinations of $b$-jets and $c$-jets from the top quark decay chain and from initial-state radiation. In order to reduce the \ttbar\ background, events are rejected if they have at least one light lepton (electron or muon) or if a hadronically decaying top quark candidate is found in the event. The top quark candidate is formed from three jets of which one must be a constituent jet of a Higgs boson candidate and is treated as the $b$-jet originating from the top decay. The other two jets form the $W$ boson from the top decay. At least one of the jets forming the $W$ boson is required not to be a constituent jet of a Higgs boson candidate since at least one of the jets from the $W$ decay must be a light jet for which the mistag probability is very low. The probablity of compatibility with the top quark decay hypothesis is then determined using the variable
 
\begin{equation*}
X_{Wt} = \sqrt{\left( \frac{m_W - 80.4\,\ \gev}{0.1 \times m_W} \right)^2 + \left( \frac{m_t - 172.5\,\ \gev}{0.1 \times m_t} \right)^2 },
\label{eqn:xwt}
\end{equation*}
 
where $m_W$ and $m_t$ are the reconstructed $W$ boson and top quark candidate masses and $0.1 \times m_W$ and $0.1 \times m_t$ are their approximate mass resolutions. If a combination of jets in the event gives $X_{Wt} < 1.8$, there is a high probability of compatibility with the top quark hypothesis and the event is vetoed. The combination of the lepton veto and the criterion for $X_{Wt}$ removes approximately 65\% of the leptonic \ttbar\ events with a signal efficiency of at least 85\%. After applying the selection, the contribution from \ttbar\ production is ~3\% of the total yield and more than 50\% for $\met > 200$ \gev.
 
The signal region (SR) is defined by the requirement
\begin{eqnarray*}
X_{hh}^\textrm{SR} &=& \sqrt{ \left( \frac{m_{2j}^\textrm{lead} - 120\ \gev}{0.1\times m_{2j}^\textrm{lead}} \right)^2 + \left( \frac{m_{2j}^\textrm{subl} - 110\ \gev}{0.1\times m_{2j}^\textrm{subl}} \right)^2} \ <\ 1.6,
\end{eqnarray*}
where $0.1 \times m_{2j}^\textrm{lead}$ and $0.1 \times m_{2j}^\textrm{subl}$ represent the approximate mass resolution of the leading and subleading Higgs boson candidates, respectively. The central values for the masses of the Higgs boson candidates of 120 \gev\ and 110 \gev, as well as the value of the $X_{hh}^\textrm{SR}$ cut, were optimized using the data-driven background model described in Section~\ref{sec:background_lowmass} and simulated signal events.
 
Additionally, as described in Section~\ref{sec:samples}, the events are required to pass at least one of three triggers requiring multiple jets or $b$-tagged jets. For signal events passing the full selection, this combination of triggers is more than 90\% efficient for the 130\,\GeV\ mass point, rising to 100\% efficiency for higgsino masses of 400\,\GeV\ and above. The per-event efficiency of this trigger combination is determined using per-jet efficiencies measured to a precision of $\sim1\%$ in dileptonic \ttbar\ events. These per-jet efficiencies are then converted to per-event efficiencies using a MC-based method that accounts for jet--jet correlations. The uncertainties in the final per-event trigger efficiencies is estimated to be $\sim2\%$.
 
Several variables are investigated to identify those most sensitive to the signal. By applying the statistical analysis described in Section~\ref{sec:results}, it is found that \met\ and \meffb\ provide the highest sensitivity. The \met is a powerful discriminant for moderate-mass higgsinos, while low-mass higgsinos are obscured by the high level of background at low \met. The variable \meffb gives better discrimination for these low-mass higgsinos. To gain from possible correlations between the two variables, the final discriminant used in the statistical analysis is the two-dimensional distribution of events in both variables via a histogram with the following lower bin edges:
\begin{eqnarray*}
\met &=& \{0, 20, 45, 70, 100, 150, 200\}\ \gev \\
\meffb &=&\{160, 200, 260, 340, 440, 560, 700, 860\}\ \gev
\end{eqnarray*}
 
In addition, two dedicated signal regions provide robust single-bin regions optimized for the discovery of SUSY signatures. The two regions are optimized using signals for the 150 \gev\ and 300 \gev\ mass points, which are representative of the mass range where this analysis is sensitive. The region definitions are given in Table \ref{tab:disc_reg_lowmass}.

\begin{table}[!t]
\begin{center}
\caption{Discovery region definitions in the low-mass analysis. The variables are defined in Section~\ref{high_event_selection}.}
\label{tab:disc_reg_lowmass}
{\footnotesize
\begin{tabular}{lcc}
\toprule
&  \multicolumn{2}{c}{Lower bound [GeV]} \\
Region                   &  \met   &   \meffb   \\
\midrule
low-SR-MET0-meff440    &  \multicolumn{1}{r}{0}    &    \multicolumn{1}{r}{440}     \\
low-SR-MET150-meff440  &  \multicolumn{1}{r}{150}  &    \multicolumn{1}{r}{440}     \\
\bottomrule
\end{tabular}
}
\end{center}
\end{table}

\subsubsection{Background estimation}\label{sec:background_lowmass}
The background is estimated using a fully data-driven method. It relies on an independent sample of events with very low signal contamination selected using the same triggers and selection criteria as described in section \ref{sec:selection_btrig} except that instead of four $b$-tagged jets, exactly two $b$-tagged jets and at least two jets that are not $b$-tagged are required. The two non-$b$-tagged jets are chosen randomly from the other jets in the event, and the two Higgs boson candidates are then formed by minimizing $D_{hh}$. The resulting sample is referred to as the ``2-tag'' sample and is approximately 200 times larger than the sample with four $b$-tagged jets, hereafter referred to as the ``4-tag'' sample.
 
The background estimate in the 4-tag sample is obtained by reweighting the events in the 2-tag sample to take into account the differences introduced by the additional $b$-tagging. These differences arise because the $b$-tagging efficiency and the $c$- and light-jet mistag rates vary as a function of jet \pt and $\eta$, the various multijet processes contribute in different proportions, and the fraction of events passed by each trigger changes.
 
To derive the background model and estimate uncertainties in the background prediction, the following regions in the mass plane of the leading and subleading \pt Higgs boson candidates are defined: control region (CR), validation region 1 (VR1) and validation region 2 (VR2), using the variables
\begin{eqnarray*}
R_{hh}^\textrm{CR} &\equiv& \sqrt{(m_{2j}^\textrm{lead} - 126.0\ \gev)^2 + (m_{2j}^\textrm{subl} - 115.5\ \gev)^2}, \label{eqn:lowmass_CR}\\
X_{hh}^\textrm{VR1} &\equiv& \sqrt{ \left( \frac{m_{2j}^\textrm{lead} - 96\ \gev}{0.1\times m_{2j}^\textrm{lead}} \right)^2 + \left( \frac{m_{2j}^\textrm{subl} - 88\ \gev}{0.1\times m_{2j}^\textrm{subl}} \right)^2 },\label{eqn:lowmass_VR1}\\
X_{hh}^\textrm{VR2} &\equiv& \sqrt{ \left( \frac{m_{2j}^\textrm{lead} - 149\ \gev}{0.1\times m_{2j}^\textrm{lead}} \right)^2 + \left( \frac{m_{2j}^\textrm{subl} - 137\ \gev}{0.1\times m_{2j}^\textrm{subl}} \right)^2 }.\label{eqn:lowmass_VR2}
\end{eqnarray*}
 
All regions satisfy the same selection criteria as those for the SR, except for the requirement on $X_{hh}^\textrm{SR}$. The control region is defined by $R_{hh}^\textrm{CR} < 55$ \gev\ and excludes the SR, $X_{hh}^\textrm{SR} > 1.6$. The two validation regions are defined by functional forms similar to that of the SR but are displaced towards lower and higher Higgs boson candidate masses satisfying $X_{hh}^\textrm{VR1} < 1.4$ and $X_{hh}^\textrm{VR2} < 1.25$, respectively. The CR center (126,115) was set so that the means of the Higgs candidates' mass distributions in the control region are equal to those in the signal region. The VR definitions were optimized to be similar to the SR while retaining sufficient statistical precision to test the background model. The CR and VRs are defined in both the 2-tag and 4-tag samples. Figure \ref{fig:btrig_massplane} shows the distributions of $m_{2j}^\textrm{lead}$ versus $m_{2j}^\textrm{subl}$ for the 2-tag and the 4-tag data after the event selection with the region definitions superimposed.
 
The background model is determined by deriving the reweighting function from the 2-tag and 4-tag data in the CR. The background estimate in the 4-tag SR is then produced from the 2-tag data in the SR by applying the reweighting function derived in the CR. The uncertainties related to the extrapolation into the SR are estimated by using the background model to reweight the 2-tag data in the validation regions and studying the differences relative to the 4-tag data in these regions. When estimating the extrapolation uncertainties, the background model is rederived while excluding the validation regions from the CR in order to obtain an unbiased estimate of the uncertainties. The uncertainties in the background model are further described in Section \ref{sec:syst}.
\begin{figure}[!t]
\centering
\subfigure[]{\includegraphics[width=0.47\textwidth]{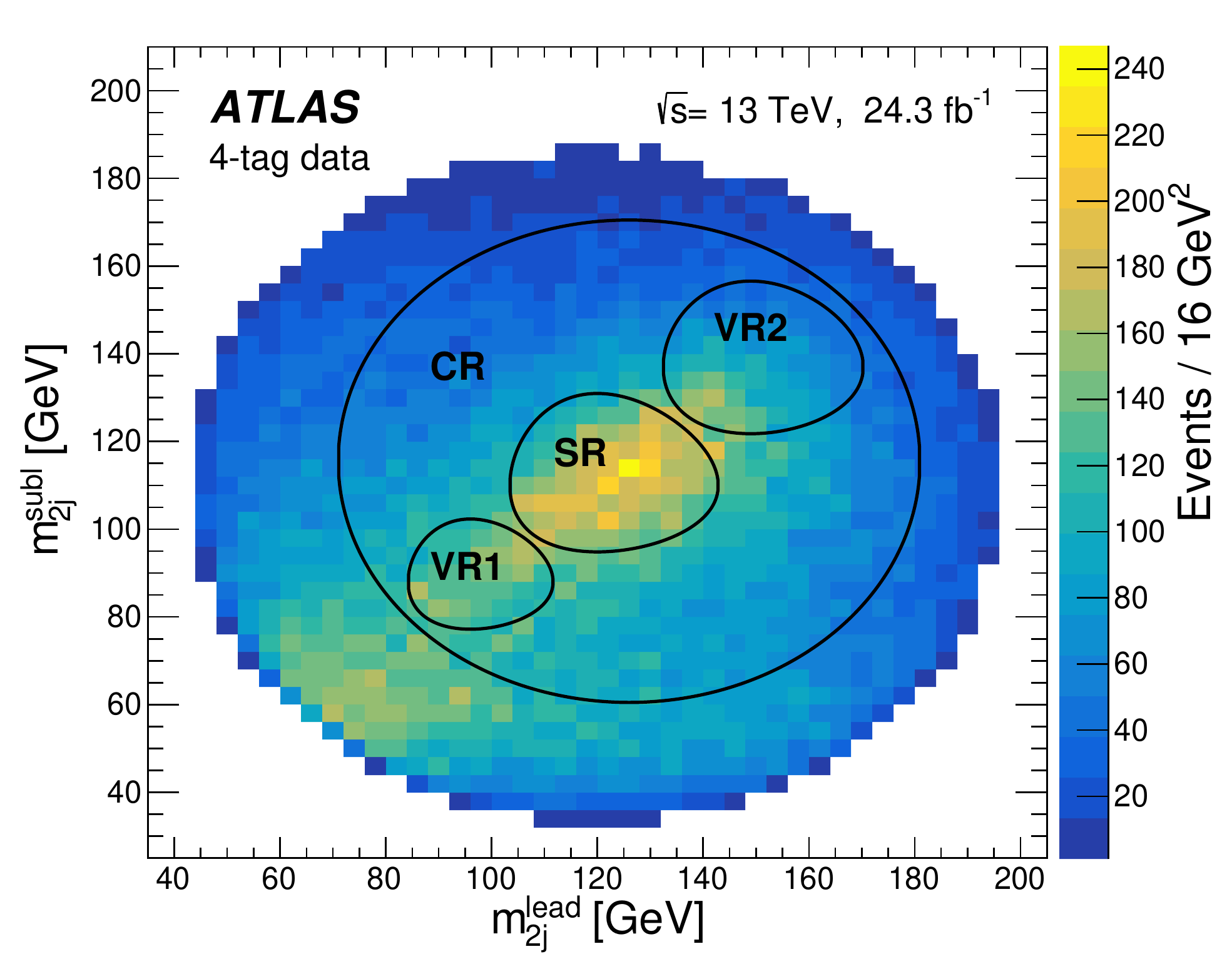}}
\subfigure[]{\includegraphics[width=0.47\textwidth]{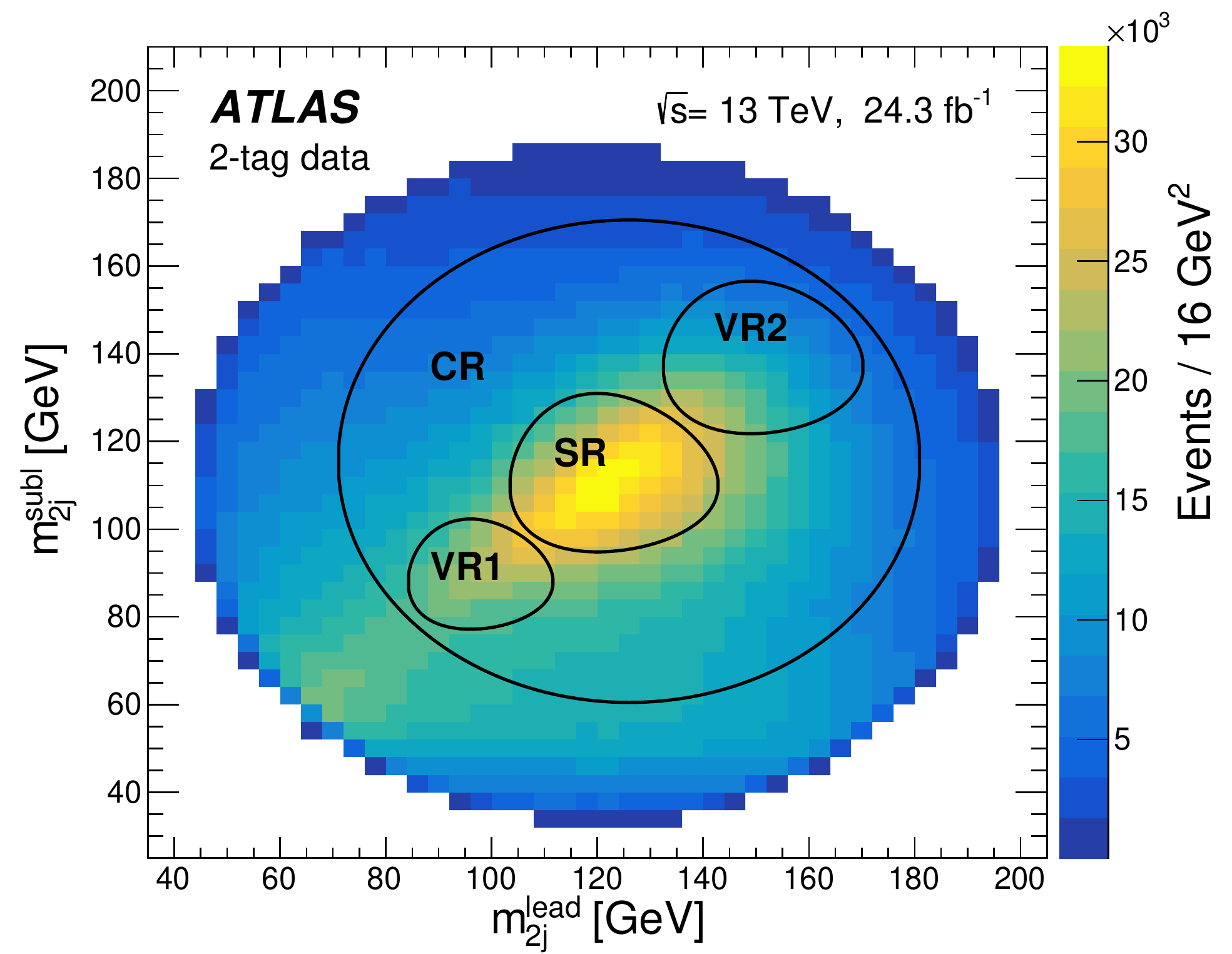}}
\caption{The distribution of $m_{2j}^\textrm{lead}$ versus $m_{2j}^\textrm{subl}$ for (a) the 4-tag data, and (b) the 2-tag data used to model the background. The region definitions are superimposed.}
\label{fig:btrig_massplane}
\end{figure}

The reweighting function defining the background model is split into an overall normalization and a component that describes the kinematical differences between the 2-tag and 4-tag data. The measured value of the normalization factor, $\mu_\textrm{2-tag}$, found in the CR is
\begin{eqnarray*}
\mu_\textrm{2-tag} &=& \frac{n_\textrm{4-tag}}{n_\textrm{2-tag}} \ =\ \left(6.03 \pm 0.03\right)\times10^{-3}.
\end{eqnarray*}
where $n_\textrm{2-tag/4-tag}$ denotes the number of 2-tag or 4-tag events, respectively, and the quoted uncertainty is the statistical uncertainty of the event yields in the CR.
 
The differing composition of 2-tag and 4-tag regions can create kinematic differences between these samples. For example, events in the 4-tag region are often produced through two gluons splitting to $b\bar{b}$ pairs, resulting in pairs of jets closer to each other, while this process contributes a smaller fraction of the 2-tag events. To correct for the kinematic differences between the 2-tag and 4-tag data, the 2-tag events from the CR are reweighted using boosted decision trees (BDT) based on the hep\_ml toolkit \cite{hep_ml}. This regression BDT allows the reweighting of events based on multiple variables simultaneously, correctly treating their correlations, while avoiding the ``curse of dimensionality'' that afflicts approaches based on multi-dimensional histograms. The BDT reweighting method was previously used by the LHCb experiment~\cite{LHCB}. 
 
At each node of the decision tree, all the input variables of the BDT are tested with requirements that split the distribution of that variable into two bins. The split that produces the two-bin distribution with the maximum $\chi^2$ between the 2-tag and 4-tag distribution is used to split the node into two sub-nodes. This process identifies the region in phase space where the difference between the 2-tag and 4-tag data is largest and therefore requires the largest correction factor. The splitting repeats for subsequent nodes of the tree, until reaching a set of stop criteria defined by the hyperparameters. The hyperparameters used in the BDT along with their values are the following: maximum number of layers (5), minimum number of events per node (250), number of trees (100), event sampling fraction (0.7), and learning rate (0.25). The BDT hyperparameters are optimized to provide a robust reweighting procedure with good statistical precision for the weights by using relatively few layers, which divide the entire space of variables into only $\mathcal{O}(30)$ regions.
 
After the tree is formed, each endpoint bin (leaf) contains a number of events for 2-tag and 4-tag data. The ratio of these, $\mu_{\textrm{leaf}} = \sum_{i} n_{\textrm{4-tag}}/ \sum_{j} n_{\textrm{2-tag}}$, is the reweighting correction for the 2-tag events on that leaf. The reweighting correction is multiplied by the learning rate, $0< \lambda \leq1$, and then applied to the 2-tag events as a scaling factor, $\exp(\lambda\log \mu_{\textrm{leaf}})$, before the procedure is repeated with the formation of a new decision tree (cf. boosting in a standard BDT for discrimination). The final weight for a given 2-tag event is the product of the weights from each individual tree, $\prod \exp(\lambda\log\mu_{\textrm{leaf}})$, renormalized to the total number of 4-tag events.

The variables passed to the reweighting BDT are optimized by identifying one at a time the single most important variable to be added to the set of variables until no further improvement in the reweighting is observed. The resulting set consists of 27 variables, including the \pt, $\eta$, and the $\Delta R$ separation of the Higgs boson candidate jets; the \pt and separation in $\eta$ of each Higgs boson candidate; the di-Higgs invariant mass; \met; $X_{Wt}$;  and information about jet multiplicity and substructure. Figure \ref{fig:lowmass_BDT_rew} shows the distribution of \met in the CR (a) before and (b) after the reweighting is applied. It is seen that the reweighted \met spectrum agrees well with the 4-tag data in the control region. The other variables used in the BDT training are also well-modeled. Figure \ref{fig:btrig_bgModel_CR} shows the background prediction from the BDT and data in the CR in the unrolled two-dimensional distribution of \met and \meffb.
 
\begin{figure}[!t]
\centering
\subfigure[]{\includegraphics[width=0.47\textwidth]{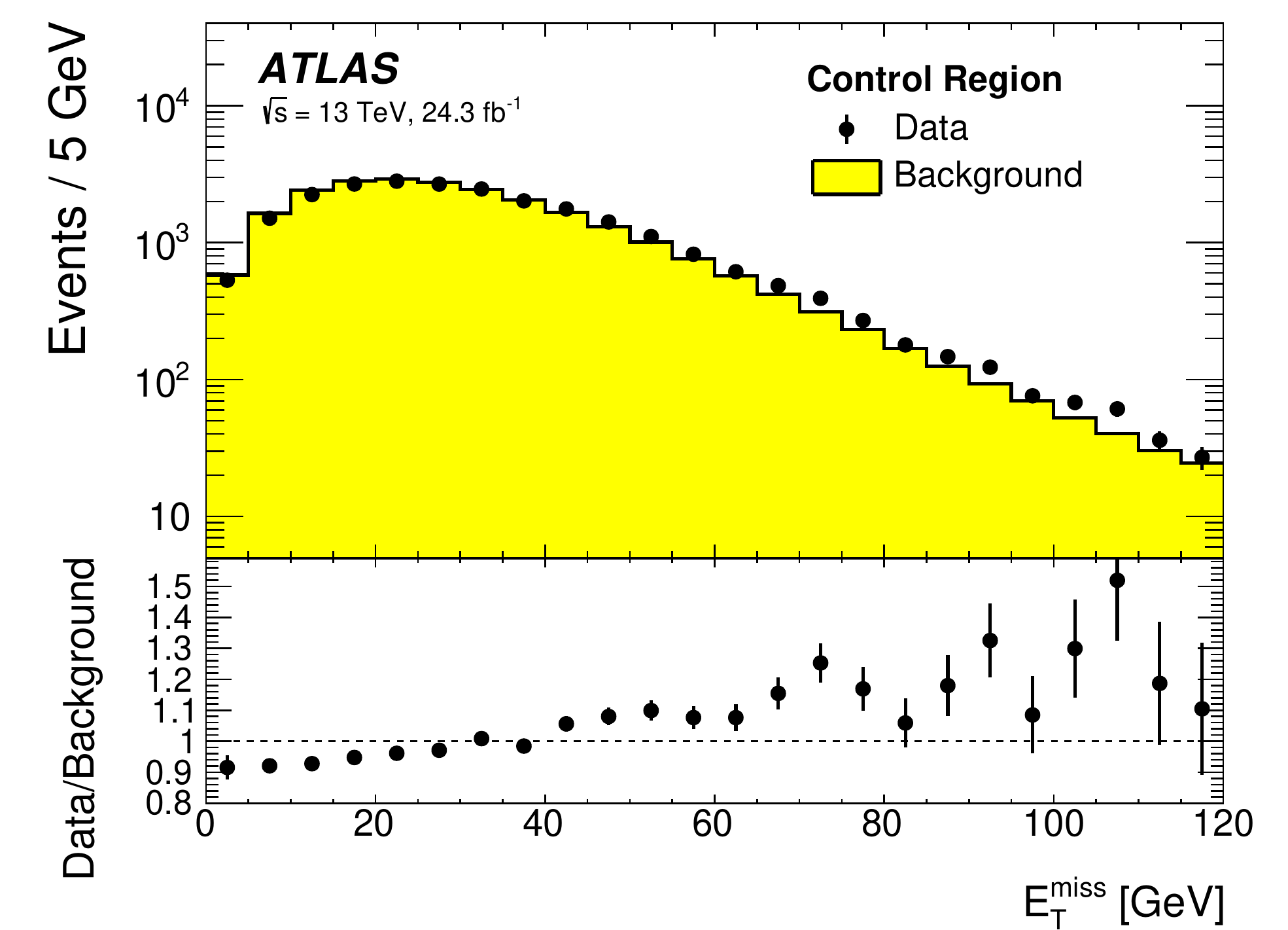}}
\subfigure[]{\includegraphics[width=0.47\textwidth]{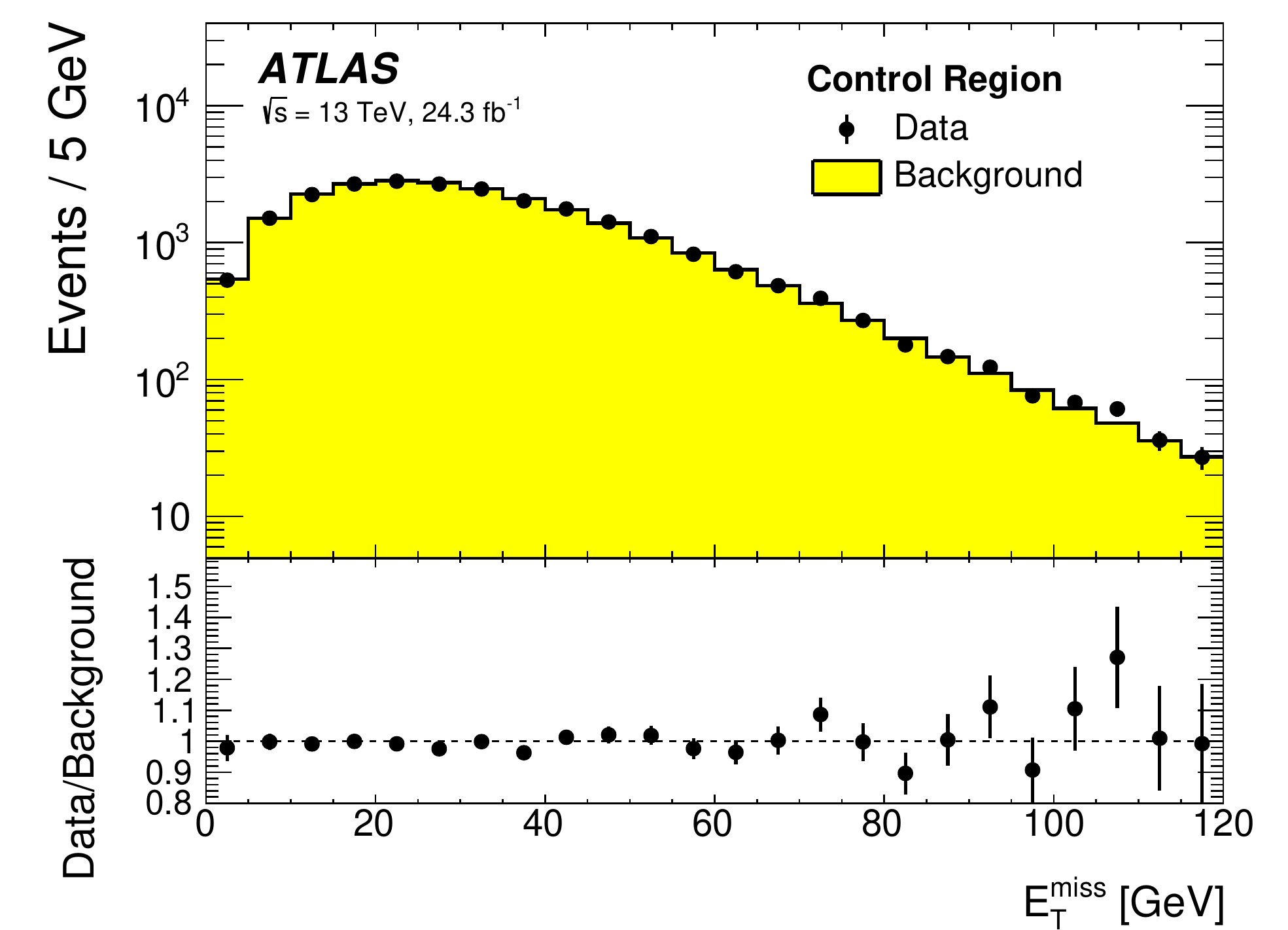}}
\caption{Distribution of \met in the control region (a) before and (b) after the BDT reweighting is applied.}
\label{fig:lowmass_BDT_rew}
\end{figure}

The background prediction is cross-checked with an alternative model where the BDT is replaced with an iterative one-dimensional reweighting method using one-dimensional projections to derive the correction factors. The correction factors are determined and applied for one variable at a time, iterating over all variables three times. This is done in a fully data-driven model and in a partially data-driven model where simulation is used to model the contributions from \ttbar\ and $Z(\rightarrow\nu\nu)$ + jets. Good agreement is found in all cross-checks.
 
\begin{figure}[!t]
\centering
\includegraphics[width=1\textwidth]{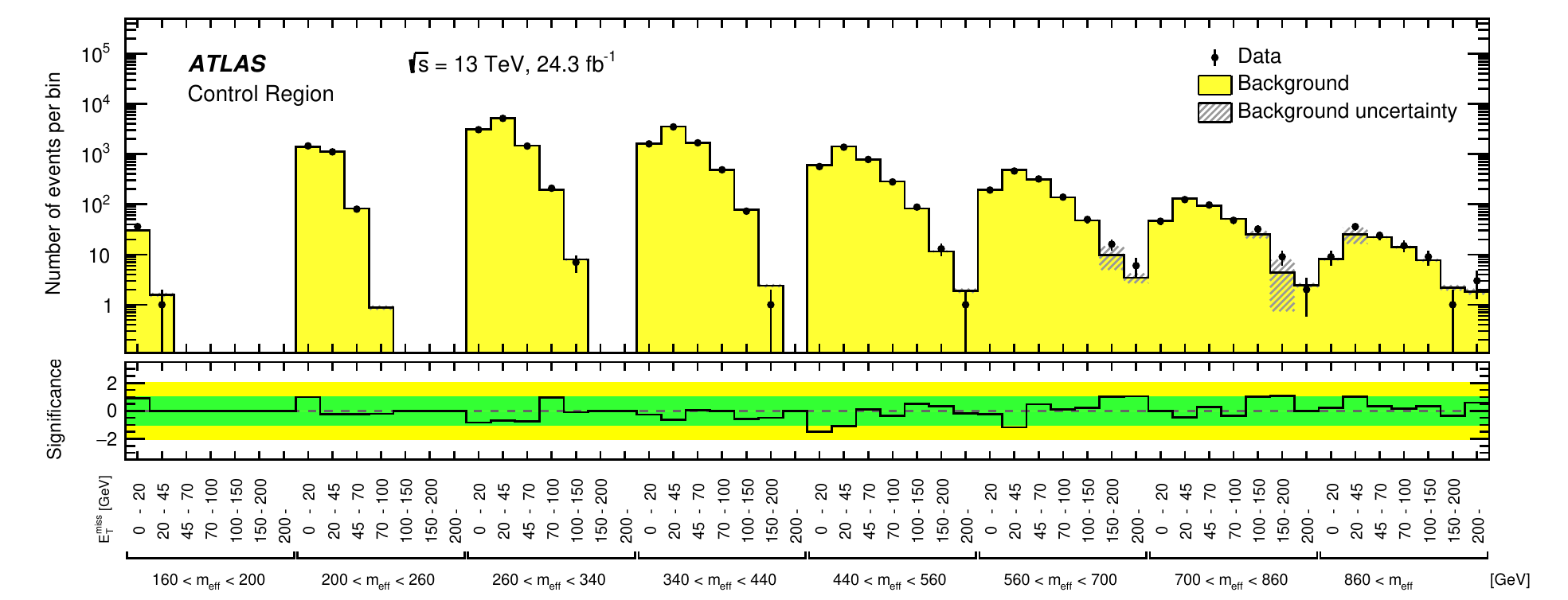}
\caption{The unrolled distribution of \met and \meffb for data and background in the control region of the low-mass analysis. The bottom panel shows the significance of any disagreement between the data and the background model. Only the statistical and non-closure uncertainties, described in Section~\ref{sec:syst_low}, are shown.}
\label{fig:btrig_bgModel_CR}
\end{figure}

 
\section{Systematic uncertainties}
\label{sec:syst}

\subsection{High-mass analysis}
\label{sec:syst_high}
 
The systematic uncertainties in the background prediction for the signal regions of the high-mass analysis arise from the extrapolation of the $\ttbar$ normalization obtained in the CRs to the SRs as well as from the yields of the minor backgrounds in the SRs, which are predicted by the simulation.

The detector-related systematic uncertainties affect both the background estimate and the signal yield.
The largest sources in this analysis relate to
the jet energy scale (JES), jet energy resolution (JER) and the $b$-tagging efficiencies and
mistagging rates. The JES uncertainties are derived from $\sqrt{s}= 13~\tev$
data and simulations~\cite{PERF-2016-04} while the JER uncertainties are extrapolated
from 8~\tev\ data using MC simulations~\cite{ATL-PHYS-PUB-2015-015}. The impact of the JES uncertainties on the expected background yields is
between 5\% and 60\%, while JER uncertainties affect the background yields by approximately 10--50\% in the
various regions. Uncertainties in the measured $b$-tagging efficiencies and mistagging rates are the subleading sources
of experimental uncertainty.
The impact of these uncertainties on the expected background
yields 
is 10--60\% depending on the region. 
All jet measurement uncertainties are propagated to the calculation
of \met, and additional uncertainties are included in the scale and resolution of the soft term. The
overall impact of the \met soft-term uncertainties is also small.

Since the normalization of the \ttbar\ background is extracted from data in the CRs, uncertainties in the
modeling of this background only affect the extrapolation from the CRs to the SRs and VRs.
Hadronization and parton shower modeling, matrix element modeling, and initial- and final-state radiation modeling are assessed by the procedures described in Ref.~\cite{ATL-PHYS-PUB-2016-004}.
An additional uncertainty is assigned to the fraction of \ttbar events produced in association with additional
heavy-flavor jets (i.e. $\ttbar\ + \geq \mathrm{1}b$ and $\ttbar\ + \geq \mathrm{1}c$), a process that has large theoretical uncertainties.
Simulation studies show that the heavy-flavor fractions in each set of SR, CR and VR, which have almost identical
$b$-tagged jets requirements, are similar.  Therefore, the theoretical uncertainties in this fraction affect these regions in
a similar way and largely cancel out in the semi-data-driven $\ttbar$ normalization based on the observed CR yields.
The residual uncertainty in the \ttbar prediction is taken as the difference between the nominal \ttbar prediction and
the one obtained after varying the cross-section of \ttbar events with
additional heavy-flavor jets by 30\%, in accordance with the results of the ATLAS measurement of this cross-section
at $\sqrt{s}= 8~\tev$~\cite{TOPQ-2014-10}. This component typically makes a small contribution (0--8\%) to the total impact of the \ttbar\ modeling
uncertainty in the background yields, which ranges between 10\% and 45\% for the various regions.
The statistical uncertainty due to the finite size of the CR samples used to extract the $\ttbar$ normalization factors, which is included in the systematic uncertainties, ranges from 5\% to 25\% depending on the SR.
 
Modeling uncertainties affecting the single-top process arise especially from the interference between the \ttbar\ and $Wt$ processes.
This uncertainty is estimated using inclusive $WWbb$ events, generated using \MGMCatNLO, which are compared with the sum of \ttbar\
and $Wt$ processes also generated with \MGMCatNLO. Radiation and parton shower modeling uncertainties are assessed as described in Ref.~\cite{ATL-PHYS-PUB-2016-004}. An additional 5\% uncertainty is included in the cross-section of
single-top processes~\cite{Kant:2014oha}.
Overall, the modeling uncertainties affecting the single-top process lead to changes of at most 11\% in the total
yields in the various regions.
Uncertainties in the $W/Z$+jets backgrounds are estimated by varying independently the scales for factorization, renormalization and resummation
by factors of 0.5 and 2. The scale used for the matching between jets originating from the matrix element and the parton shower is also varied. The resulting
uncertainties in the total yield range from approximately 5\% to 20\% in the various regions.
A 50\% normalization uncertainty is assigned to $\ttbar+W/Z/h$, $\ttbar\ttbar$, and diboson backgrounds; this has no significant impact
on the sensitivity of this analysis.
Uncertainties arising from variations of the parton distribution functions are found to affect background yields by less than 2\%, and
therefore these uncertainties are neglected here.
Uncertainties due to the number of events in the MC background samples reach approximately
50\% in one region, but are typically 20\%.
 
Figure \ref{fig:syst_etmiss} summarizes the relative systematic uncertainties in the background estimate.
The total systematic uncertainties range from approximately 30\% to 80\% in the various SRs.
 
\begin{figure}[htbp]
\centering
\includegraphics[width=0.65\textwidth]{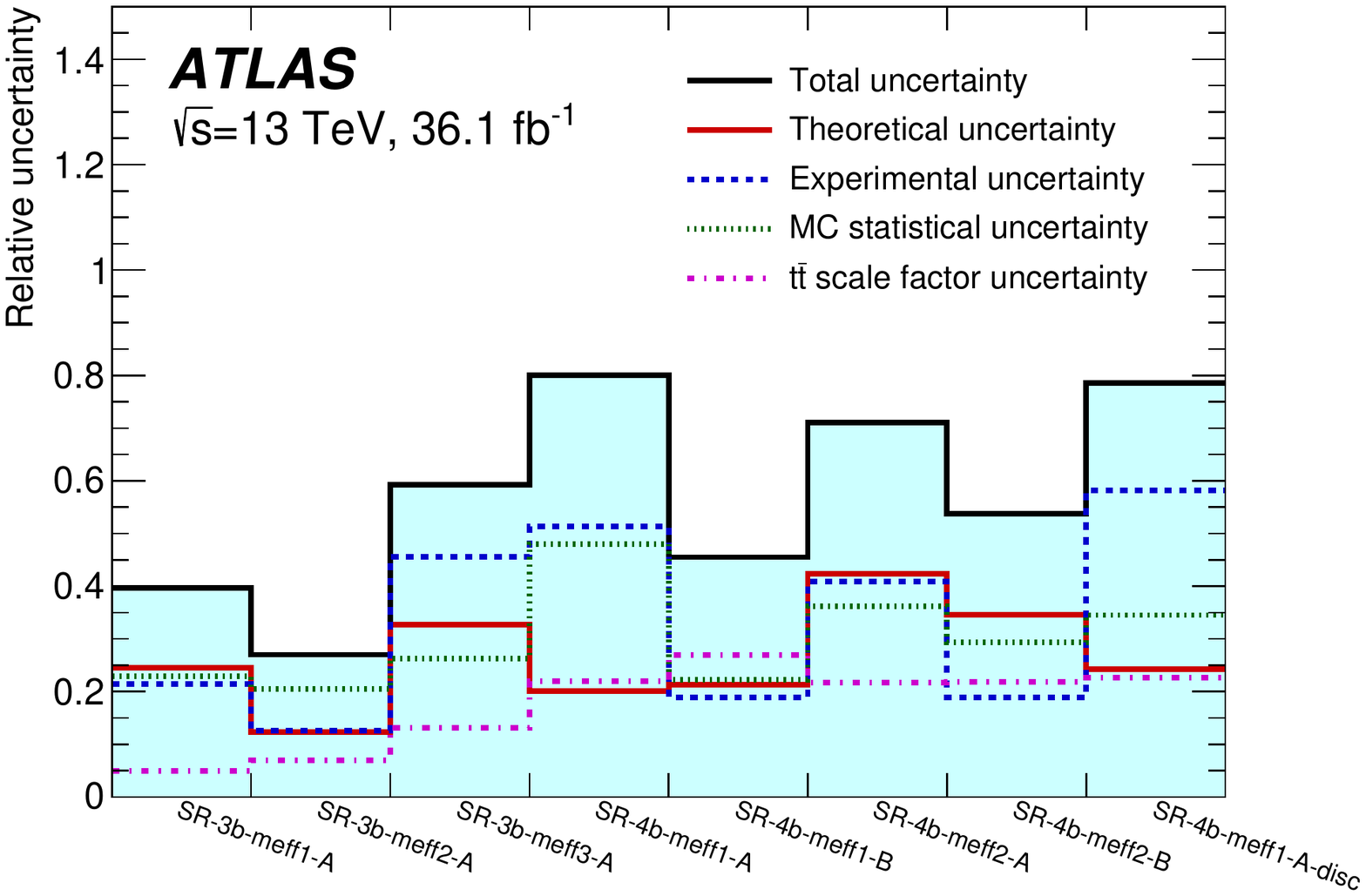}
\caption{Relative systematic uncertainties in the background estimate for the high-mass analysis. The individual uncertainties can be correlated, such that the total background uncertainty is not necessarily their sum in quadrature.
}
\label{fig:syst_etmiss}
\end{figure}
 
The uncertainties in the cross-sections of signal processes are determined from an envelope of different
cross-section predictions, as described in Section~\ref{sec:samples}. These are also applied in the low-mass analysis.

 
\subsection{Low-mass analysis}
\label{sec:syst_low}
The total uncertainty in the background prediction in the low-mass signal region has three sources:
\begin{enumerate}
\item Non-closure of the shape in the control region.
\item Validity of transfer of weights across regions.
\item Statistical uncertainty of the 2-tag data in the signal region.
\end{enumerate}
 
The non-closure uncertainty reflects any imperfections in the modeling when comparing reweighted 2-tag data to 4-tag data in the control region, which could be the result of an insufficiently flexible reweighting function that is not capable of fully correcting the 2-tag data or relevant variables not being utilized in the reweighting. The normalization of the background model is be correct by construction in the control region, but the distributions of variables are not.
 
Non-closure uncertainties are evaluated bin-by-bin by computing the difference between the data and the predicted background in the control region defined in Section \ref{sec:background_lowmass} and shown in Figure~\ref{fig:btrig_bgModel_CR}. If the difference is larger than the combined statistical uncertainty of the data and background, a non-closure uncertainty equal to the observed discrepancy is assigned to this bin. If the difference is smaller, no non-closure uncertainty is assigned. These uncertainties are treated as uncorrelated bin-to-bin in the final statistical analysis. Adding bin-to-bin correlations has no significant impact on the final results.
 
The two validation regions defined in Section \ref{sec:background_lowmass} are used to assess the validity of weight transfer across the Higgs boson candidate mass plane. To replicate the situation in the signal region as closely as possible, the background model is derived using the data in the control region and excluding the data from the validation region under study. It is verified that the background models derived with or without the data in one of the two validation regions are consistent within the uncertainties on the samples.
 
The normalization in VR1 is incorrect by 2.1\%, while in VR2 the bias is 4.0\%. The 4.0\% value is assigned as the transfer normalization uncertainty. Similarly to the non-closure uncertainty, the difference in each bin in both VR1 and VR2 is calculated after normalizing to the total yield in data. For a given bin, the larger of the two differences is assigned as the transfer shape uncertainty if the difference is larger than the combined statistical uncertainty of the data and the background. If the difference is smaller, no transfer shape uncertainty is assigned.
 
Finally, the uncertainties related to the statistical precision of the 2-tag sample are included. Figure \ref{fig:btrig_bgModel_unc} shows the different components of the background modeling uncertainty.
 
The detector modeling systematic uncertainties only affect the signal models because the background model is entirely data-driven. The detector-related systematic uncertainties include the jet energy scale and resolution, the \met soft term, and the $b$-tagging efficiency. The lepton energy scale and efficiency uncertainties are negligible given their small size and the rarity of leptons in the signal events. All detector modeling uncertainties are subdominant to the data-driven uncertainties.

\begin{figure}[!t]
\centering
\includegraphics[width=1\textwidth]{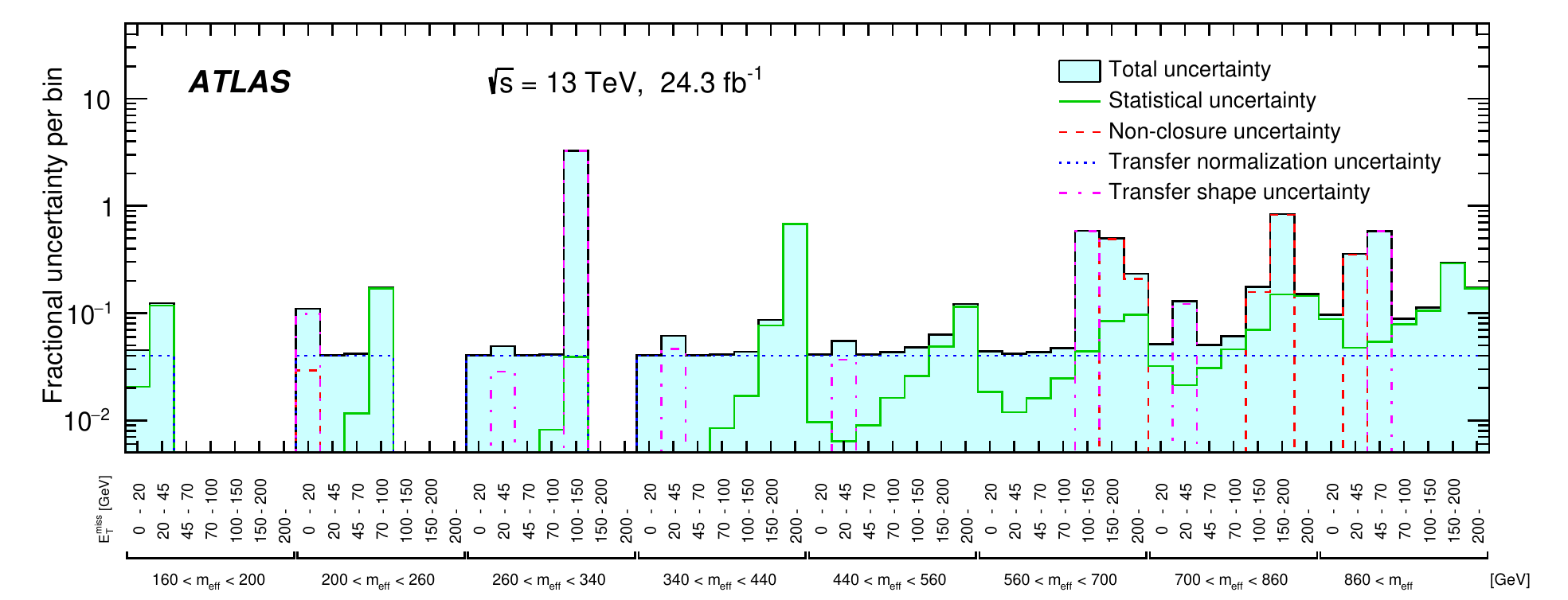}
\caption{Breakdown of relative uncertainties in background model in the low-mass analysis. Uncertainties below 0.5\% are not shown but are used in the fit.}
\label{fig:btrig_bgModel_unc}
\end{figure}
 
\section{Results}
\label{sec:results}
 
\subsection{High-mass analysis}

Figure \ref{fig:pullVR} shows the results of the
background-only fit to the CRs, extrapolated to the VRs.
The number of events predicted by the background-only fit is compared to the data in
the upper panel. The significance is the difference between the observed number of
events  and the predicted background yield
divided by the total uncertainty and is shown for each region in the
lower panel. No evidence of significant background mismodeling is observed in the
VRs.

\begin{figure}[htbp]
\centering
\includegraphics[width=0.9\textwidth]{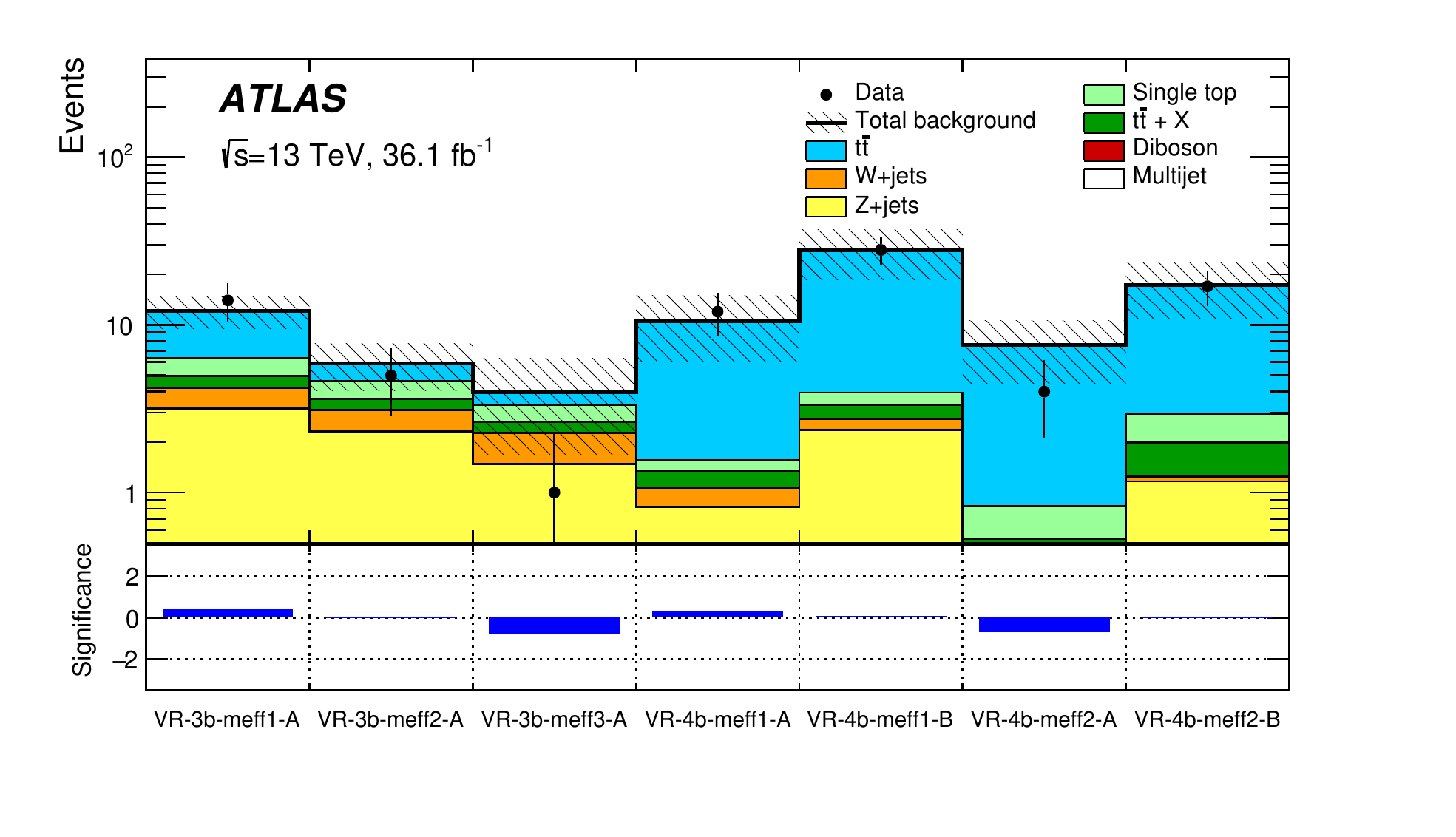}
\caption{Results of the background-only fit extrapolated to the VRs of the high-mass analysis. The $\ttbar$ normalization
is obtained from the fit to the CRs shown in Figure~\ref{fig:pullCR}. The upper panel shows
the observed number of events and the predicted background yield. The bottom panel shows the significance of any disagreement between the data and the background model~\cite{Choudalakis2012}.
All uncertainties  defined in Section~\ref{sec:syst_high} are included in the
uncertainty band. The background category $\ttbar+X$ includes $\ttbar W/Z$,
$\ttbar H$, and $\ttbar \ttbar$ events. }
\label{fig:pullVR}
\end{figure}
 
The event yields in the SRs of the high-mass analysis are presented in Figure~\ref{fig:pullSR}. The significance is shown for each region in the lower panel.   No significant excess is found above the predicted background. The background
is dominated by $\ttbar$ events in all SRs. The subdominant background contributions are
$Z(\to \nu\nu)$+jets and $W(\to \ell \nu)$+jets events, where for $W$+jets events the lepton is an unidentified electron or muon or
a hadronically decaying $\tau$-lepton. These yields are also shown in Table~\ref{tab:yieldsSR}. 
 
\begin{figure}[htbp]
\centering
\includegraphics[width=0.9\textwidth]{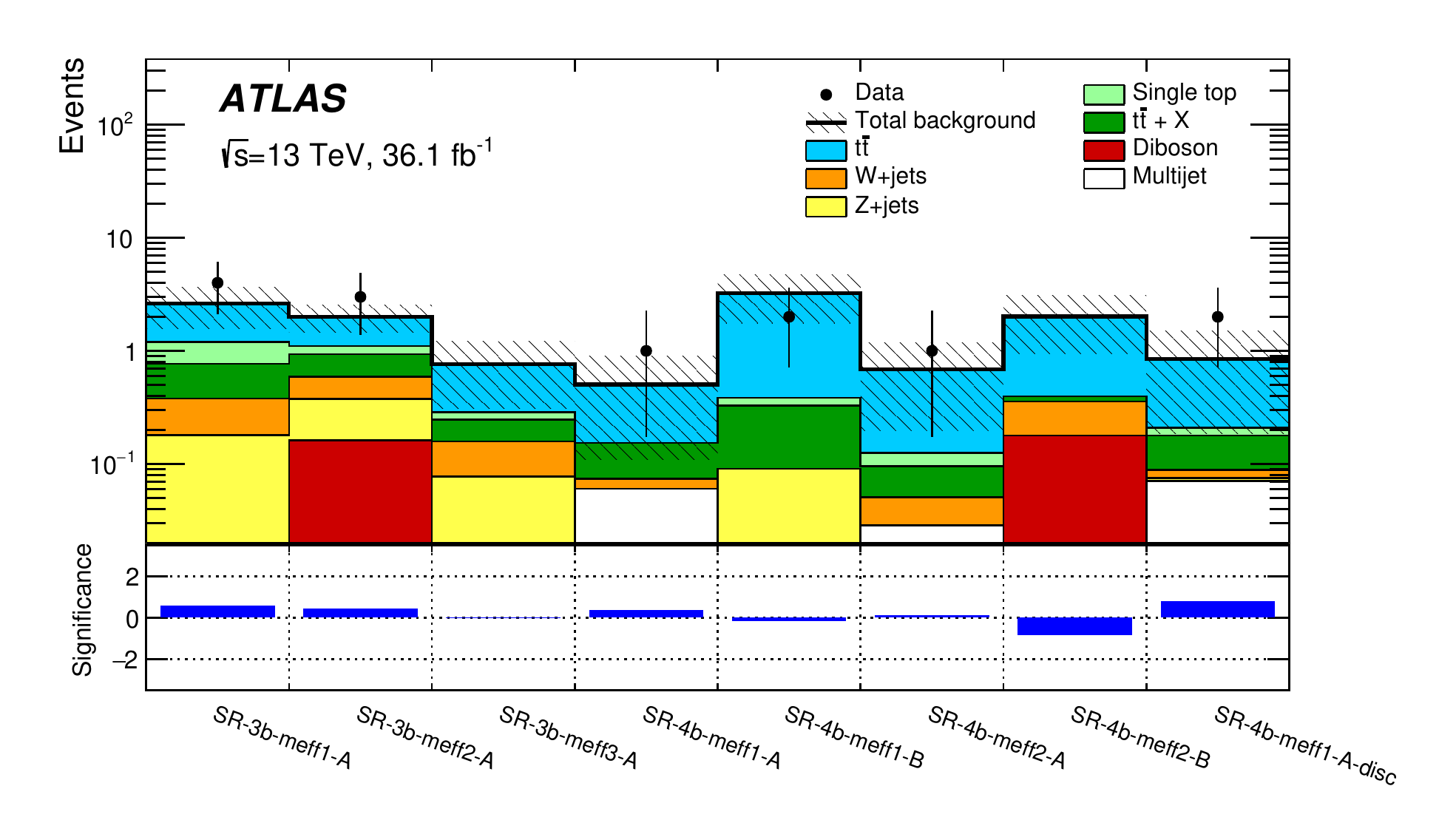}
\caption{Results of the background only fit extrapolated to the SRs of the high-mass analysis. The $\ttbar$ normalization
is obtained from the fit to the CRs shown in Figure~\ref{fig:pullCR}. The data in the  SRs are
not included in the fit.  The upper panel shows the observed number of events and the predicted background
yield.  The bottom panel shows the significance of any disagreement between the data and the background model~\cite{Choudalakis2012}. All uncertainties  defined in Section~\ref{sec:syst_high} are included in the uncertainty band. The background
category $\ttbar+X$ includes $\ttbar W/Z$, $\ttbar H$, and $\ttbar \ttbar$ events.}
\label{fig:pullSR}
\end{figure}
 
\begin{table}
\caption{Results of the background-only fit extrapolated to the SRs of the high-mass analysis, for the total background prediction and breakdown of the main background sources.
The uncertainties shown include all systematic uncertainties. The data in the SRs are not included in the fit.
The background category $\ttbar+X$ includes $\ttbar W/Z$, $\ttbar H$, and $\ttbar \ttbar$ events.
The row ``MC-only background'' provides the total background prediction when the
$\ttbar$ normalization is obtained from a theoretical
calculation~\cite{Czakon:2011xx}.}
\label{tab:yieldsSR}
\resizebox{1.\textwidth}{!}{
\begin{tabular}{|l|c|c|c|c|c|c|c|c|}
\hline
SR name & SR-3b-meff1-A & SR-3b-meff2-A & SR-3b-meff3-A & SR-4b-meff1-A & SR-4b-meff1-B & SR-4b-meff2-A & SR-4b-meff2-B & SR-4b-meff1-A-disc\\
\hline
$N_{\mathrm{obs}}$ & 4 & 3 & 0 & 1 & 2 & 1 & 0 & 2\\
\hline
Total background & 2.6 $\pm$ 1.0 & 2.0 $\pm$ 0.5 & 0.8 $\pm$ 0.5 & 0.5 $\pm$ 0.4 & 3.2 $\pm$ 1.5 & 0.7 $\pm$ 0.5 & 2.0 $\pm$ 1.1 & 0.8 $\pm$ 0.7\\
Fitted \ttbar & 1.4 $\pm$ 0.8 & 0.89 $\pm$ 0.32 & 0.5 $\pm$ 0.4 & 0.35 $\pm$ 0.33 & 2.8 $\pm$ 1.5 & 0.6 $\pm$ 0.5 & 1.6 $\pm$ 1.0 & 0.6 $\pm$ 0.6\\
Single top & 0.43 $\pm$ 0.29 & 0.17 $\pm$ 0.14 & 0.040 $\pm$ 0.017 & $<$ 0.01 & 0.06 $\pm$ 0.13 & 0.030 $\pm$ 0.019 & $<$ 0.01 & 0.030 $\pm$ 0.019\\
$\ttbar+X$ & 0.39 $\pm$ 0.16 & 0.34 $\pm$ 0.14 & 0.09 $\pm$ 0.04 & 0.08 $\pm$ 0.06 & 0.24 $\pm$ 0.10 & 0.045 $\pm$ 0.025 & 0.039 $\pm$ 0.033 & 0.09 $\pm$ 0.06\\
$Z$+jets & 0.18 $\pm$ 0.14 & 0.21 $\pm$ 0.16 & 0.07 $\pm$ 0.20 & $<$ 0.01 & 0.09 $\pm$ 0.04 & $<$ 0.01 & $<$ 0.01 & 0.004 $\pm$ 0.011\\
$W$+jets & 0.20 $\pm$ 0.06 & 0.21 $\pm$ 0.09 & 0.08 $\pm$ 0.06 & 0.013 $\pm$ 0.009 & $<$ 0.01 & 0.022 $\pm$ 0.027 & 0.18 $\pm$ 0.10 & 0.013 $\pm$ 0.008\\
Diboson & $<$ 0.01 & 0.16 $\pm$ 0.11 & $<$ 0.01 & $<$ 0.01 & $<$ 0.01 & $<$ 0.01 & 0.17 $\pm$ 0.08 & $<$ 0.01\\
Multijet & $<$ 0.01 & 0.004 $\pm$ 0.005 & 0.004 $\pm$ 0.006 & 0.06 $\pm$ 0.05 & 0.0027 $\pm$ 0.0021 & 0.03 $\pm$ 0.04 & 0.007 $\pm$ 0.012 & 0.07 $\pm$ 0.05\\
\hline
MC-only background & 2.5 $\pm$ 1.0 & 2.0 $\pm$ 0.5 & 0.6 $\pm$ 0.4 & 0.43 $\pm$ 0.31 & 2.6 $\pm$ 0.9 & 0.43 $\pm$ 0.27 & 1.3 $\pm$ 0.6 & 0.7 $\pm$ 0.5\\
\hline
\end{tabular}
}
\end{table}

 
\subsection{Low-mass analysis}
The unrolled two-dimensional distributions of \met and \meffb in the two validation regions for the low-mass analysis are shown Figures~\ref{fig:btrig_bgModel_VR1} and \ref{fig:btrig_bgModel_VR2}. The significances are shown in the lower panel. No significant mismodeling is observed.
 
\begin{figure}[!h]
\centering
\includegraphics[width=1\textwidth]{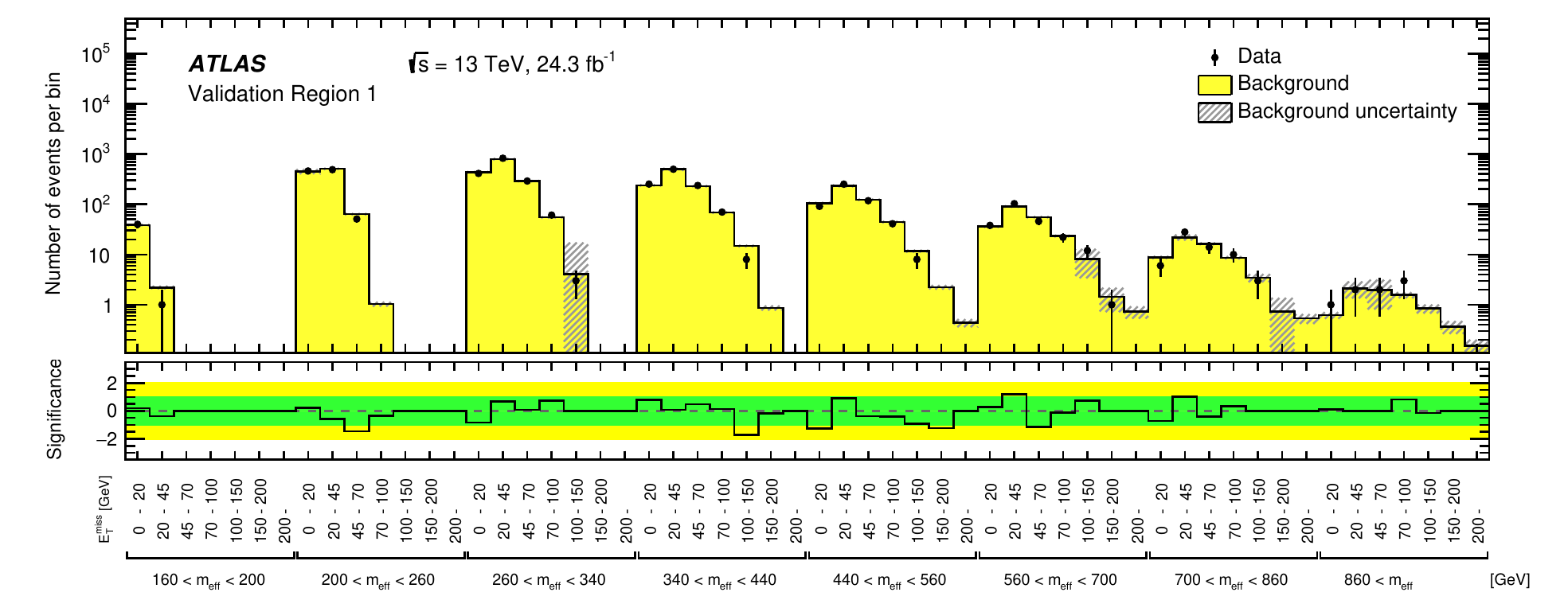}
\caption{The unrolled distribution of \met and \meffb for data and background in validation region 1 of the low-mass analysis. The bottom panel shows the significance of any disagreement between the data and the background model~\cite{Choudalakis2012}. All systematic uncertainties described in Section~\ref{sec:syst_low} are included.}
\label{fig:btrig_bgModel_VR1}
\end{figure}
 
\begin{figure}[!h]
\centering
\includegraphics[width=1\textwidth]{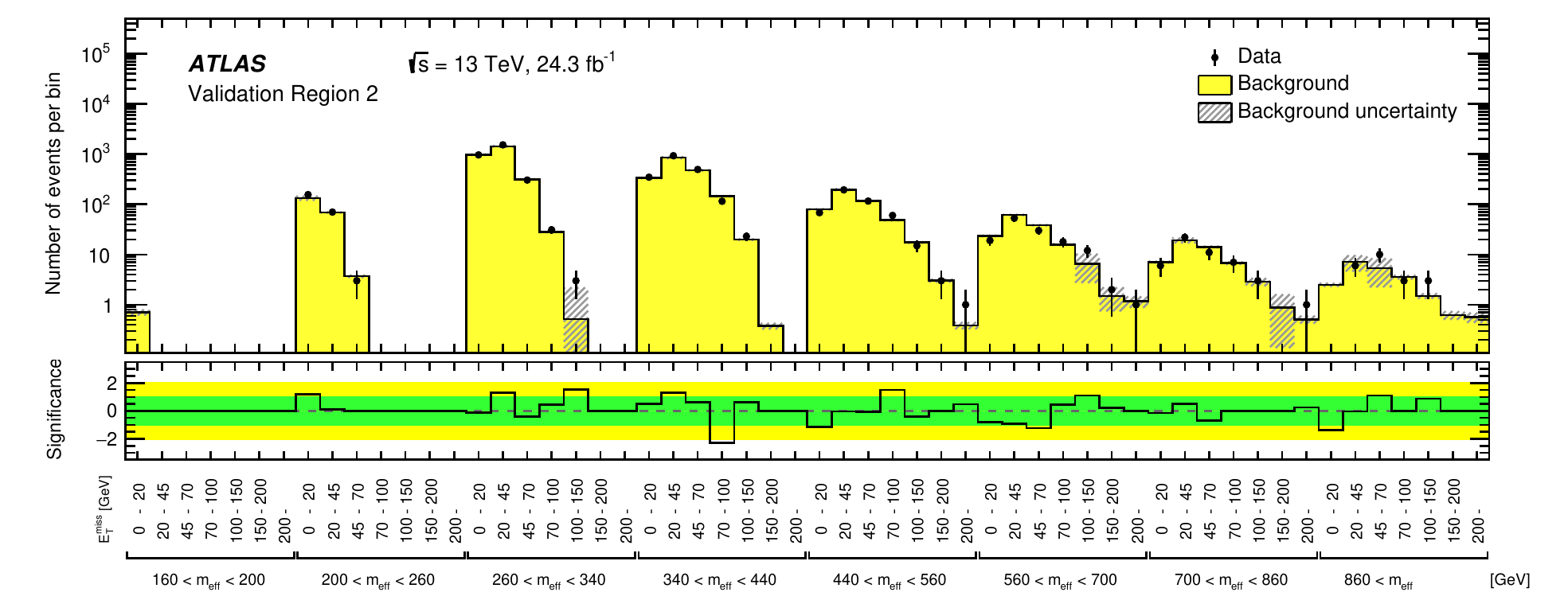}
\caption{The unrolled distribution of \met and \meffb for data and background in validation region 2 of the low-mass analysis. The bottom panel shows the significance of any disagreement between the data and the background model~\cite{Choudalakis2012}. All systematic uncertainties described in Section~\ref{sec:syst_low} are included.}
\label{fig:btrig_bgModel_VR2}
\end{figure}
 
The signal regions for the low-mass analysis are presented in Figure \ref{fig:btrig_bgModel_SR}, and the significance of any disagreement between the data and background model is shown in the bottom panel. No significant excess is found above the predicted background. The most significant upward deviation is observed in the bin $860 < \meffb < 2000$ GeV and $150 < \met < 200$ GeV, where four events are observed compared to $1.0\pm 0.2$ expected. A few other bins at high \met\ have excesses below $2\sigma$ in significance.

\begin{figure}[!t]
\centering
\includegraphics[width=1\textwidth]{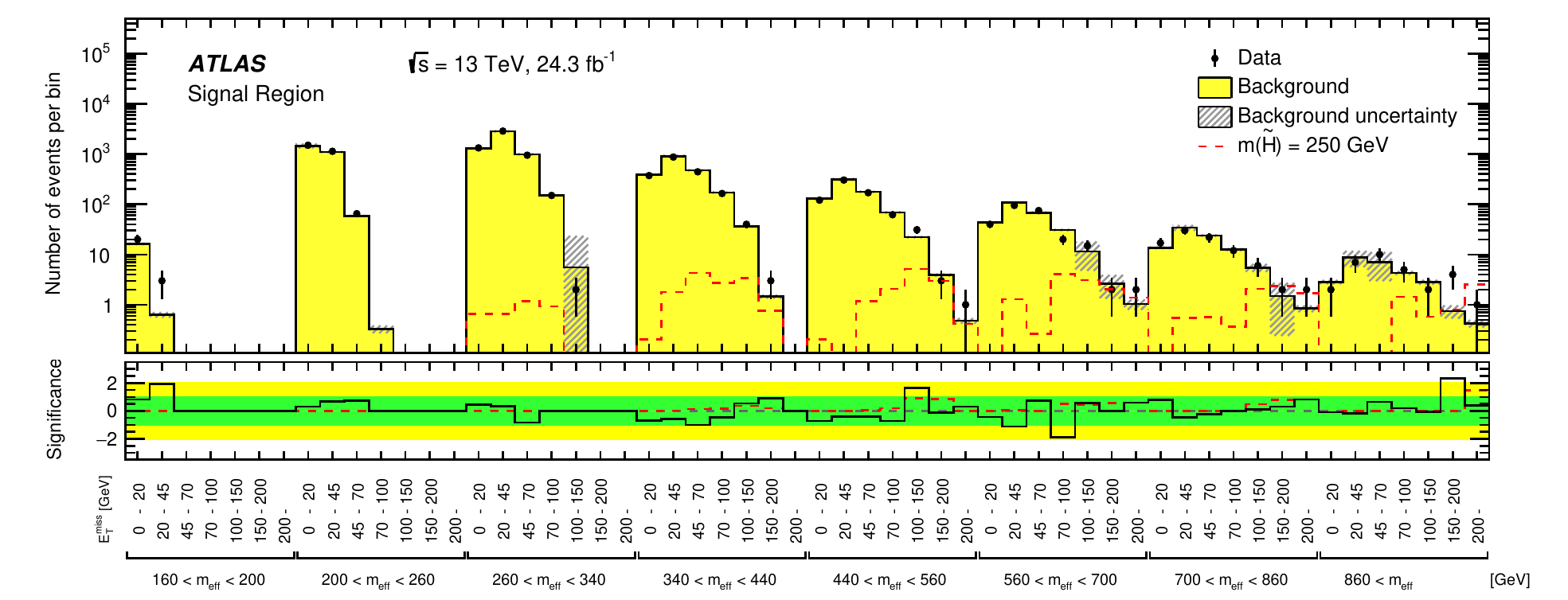}
\caption{The unrolled distribution of \met and \meffb for data, background and an example signal sample in the signal region of the low-mass analysis. The bottom panel shows the significance of any disagreement between the data and the background model~\cite{Choudalakis2012}. All systematic uncertainties described in Section~\ref{sec:syst_low} are included. The dashed line includes the signal contribution and defines the significance as signal/$\sigma$.}
\label{fig:btrig_bgModel_SR}
\end{figure}

\section{Interpretation}
\label{sec:interpretation}
Since no significant excess over the expected background from SM processes is observed, the data are used to derive one-sided upper limits at 95\% confidence level (CL). Two types of interpretation are given in this paper: model-independent exclusion
limits and model-dependent exclusion limits on degenerate \hino\ production.
 
\subsection{Model-independent exclusion limits}
 
Model-independent limits on the number of beyond-the-SM (BSM) events for each of the discovery SRs are derived with
pseudoexperiments using the CL$_\mathrm{s}$ prescription \cite{Read:2002hq} and neglecting a possible signal contamination
in the CR. Only the discovery regions from both the high-mass and low-mass analyses are used in order to simplify
the reintepretation of these limits. Limits are obtained with a fit in each SR which proceeds in the same way
as the fit used to predict the background, except that the number of events observed in the SR is included as an input to
the fit. Also, an additional parameter for the BSM signal strength,
constrained to be non-negative, is fit.
Upper limits on the
visible BSM cross-section  ($\sigma^{95}_\mathrm{vis}$) are obtained
by dividing the observed upper limits on the number of BSM events by the
integrated luminosity.
The results are given in Table~\ref{tab:UL_toys}, along with the $p_0$-values,
the probability of the SM background alone to fluctuate to the observed number of events or higher.
\begin{table}
\begin{center}
\caption[Model independent upper limits]{For each discovery region, the number of observed events ($N_\mathrm{obs}$), the number of predicted events ($N_\mathrm{pred}$), and 95\% CL upper limits on the visible cross-section ($\sigma^\mathrm{95}_\mathrm{vis}$) and on the number of signal events ($S_\mathrm{obs}^\mathrm{95}$ ) are shown.  The fifth column ($S_\mathrm{exp}^\mathrm{95}$) shows the 95\% CL upper limit on the number of signal events given the expected number (and $\pm 1\sigma$ excursions of the expectation) of background events. The last column indicates the discovery $p$-value ($p(s~=~0)$) in significance units. The $p$-values are capped at 0.5. Results are obtained with $20\,000$ pseudoexperiments.}
\label{tab:UL_toys}
\begin{tabular}{
lr
S[table-format=4.1(1)]
S[table-format=1.1(2)]
S[table-format=2.1(1)]
cc
}
\toprule
{ Signal channel}           &   $N_\mathrm{obs}$ & \multicolumn{1}{c}{$N_\mathrm{pred}$}       & \multicolumn{1}{c}{$\sigma^\mathrm{95}_\mathrm{vis}$ [fb]}  &  $S_\mathrm{obs}^\mathrm{95}$  & $S_\mathrm{exp}^\mathrm{95}$ & $p_0$ (Z)  \\
\midrule
high-SR-4b-meff1-A-disc   &    2 &     0.8 \pm 0.7  & 0.15 &   5.5 & ${ 4.2 }^{ +1.3 }_{ -0.4 }$  &  0.15$~$(1.02) \\%
high-SR-3b-meff3-A        &    0 &     0.8 \pm 0.5  & 0.08 &   3.0 & ${ 3.1 }^{ +1.2 }_{ -0.1 }$  &  0.50$~$(0.00) \\%
low-SR-MET0-meff440       & 1063 &    1100 \pm 25   & 2.3  &  56   & ${ 79 }^{ +31 }_{ -23 }$     &  0.50$~$(0.00) \\%
low-SR-MET150-meff440     &   17 &      12 \pm 8    & 0.90 &  22   & ${ 19 }^{ +5 }_{ -4 }$       &  0.21$~$(0.80) \\%
\bottomrule
\end{tabular}
\end{center}
\end{table}

\subsection{Model-dependent exclusion limits}
The results are used to place exclusion limits on the higgsino pair production signal model. The results are obtained using the
CL$_\mathrm{s}$ prescription
in the asymptotic approximation \cite{Cowan:2010js}.
The signal contamination in the CRs
and the experimental systematic uncertainties in the signal are taken into account. All of the regions of the high-mass and low-mass analyses are combined in the respective fits. The analysis with the better expected limit at each generated \hino\ mass point is selected for the combined result. The transition between the two analyses occurs at $\mhino = 300$ GeV. The results for a branching ratio for decays $\hino \rightarrow h \gravino$ of 100\% are shown in Figure~\ref{fig:exclusion_combined}. Degenerate higgsino masses between 130 GeV and 230 GeV and between 290 GeV and 880 GeV are excluded at 95\% confidence level.  In the range approximately $200\ \gev < \mhino < 300\ \gev$, the observed limit is 1--2 $\sigma$ weaker than expected, due to the data exceeding the background in several bins with \met\ $>$ 100 \gev\ in the low-mass analysis.
 
The results are also interpreted in the context of a variable branching ratio, where the \hino\ is allowed to decay to $Z$ or Higgs bosons. As with the 100\% $\hino\rightarrow h \gravino$ interpretation, the results of the low-mass analysis are used below $\mhino = 300$ GeV, while those of the high-mass analysis are used above. The combined limits are shown in Figure~\ref{fig:exclusion_br}: branching ratios for decays $\hino \rightarrow h \gravino$ as low as 45\% are excluded for $\mhino \approx 400$ \gev\ at 95\% confidence level.

\begin{figure}[htbp]
\centering
\subfigure[]{\includegraphics[width=0.49\textwidth]{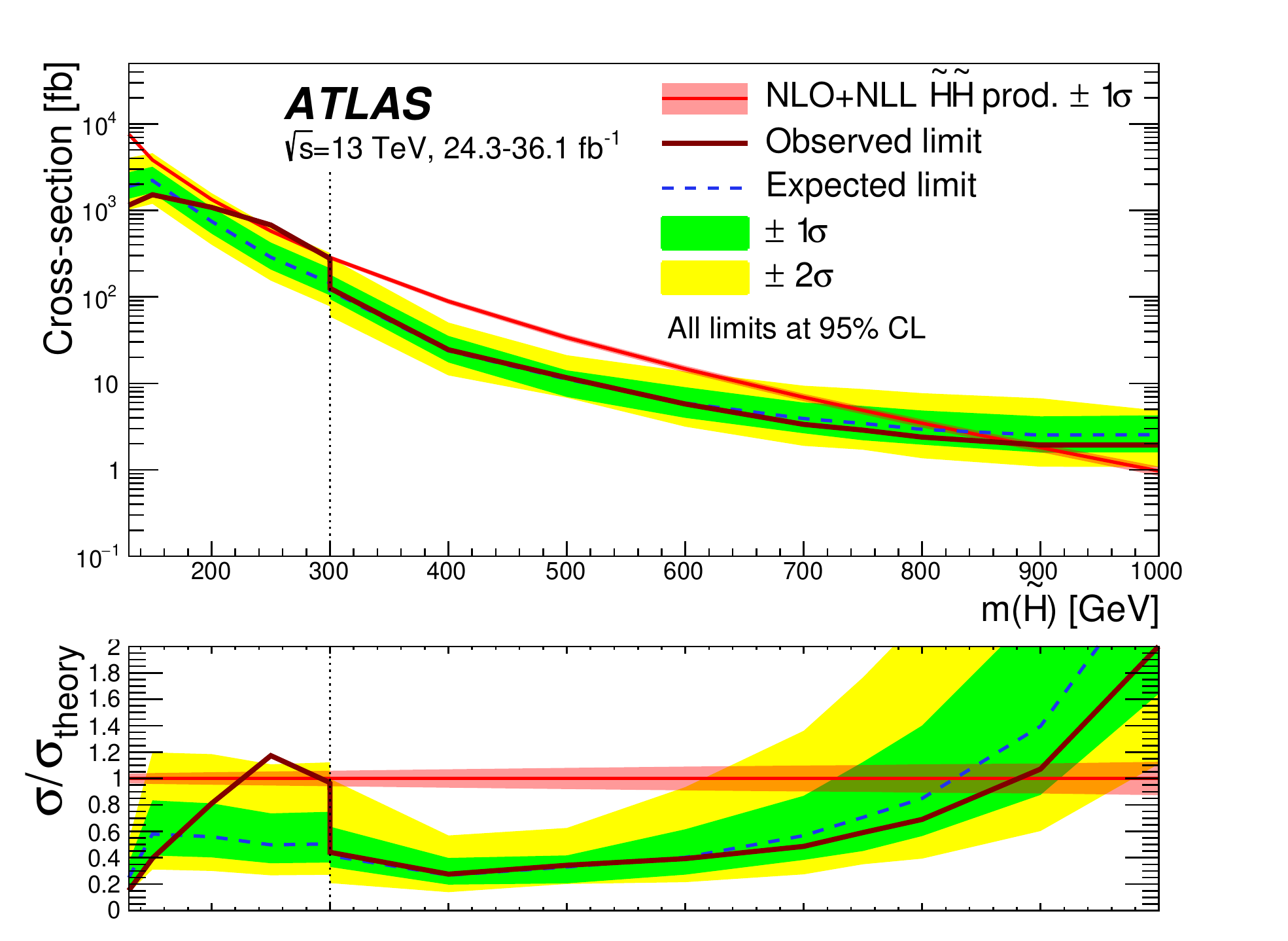}\label{fig:exclusion_combined}}
\subfigure[]{\includegraphics[width=0.49\textwidth]{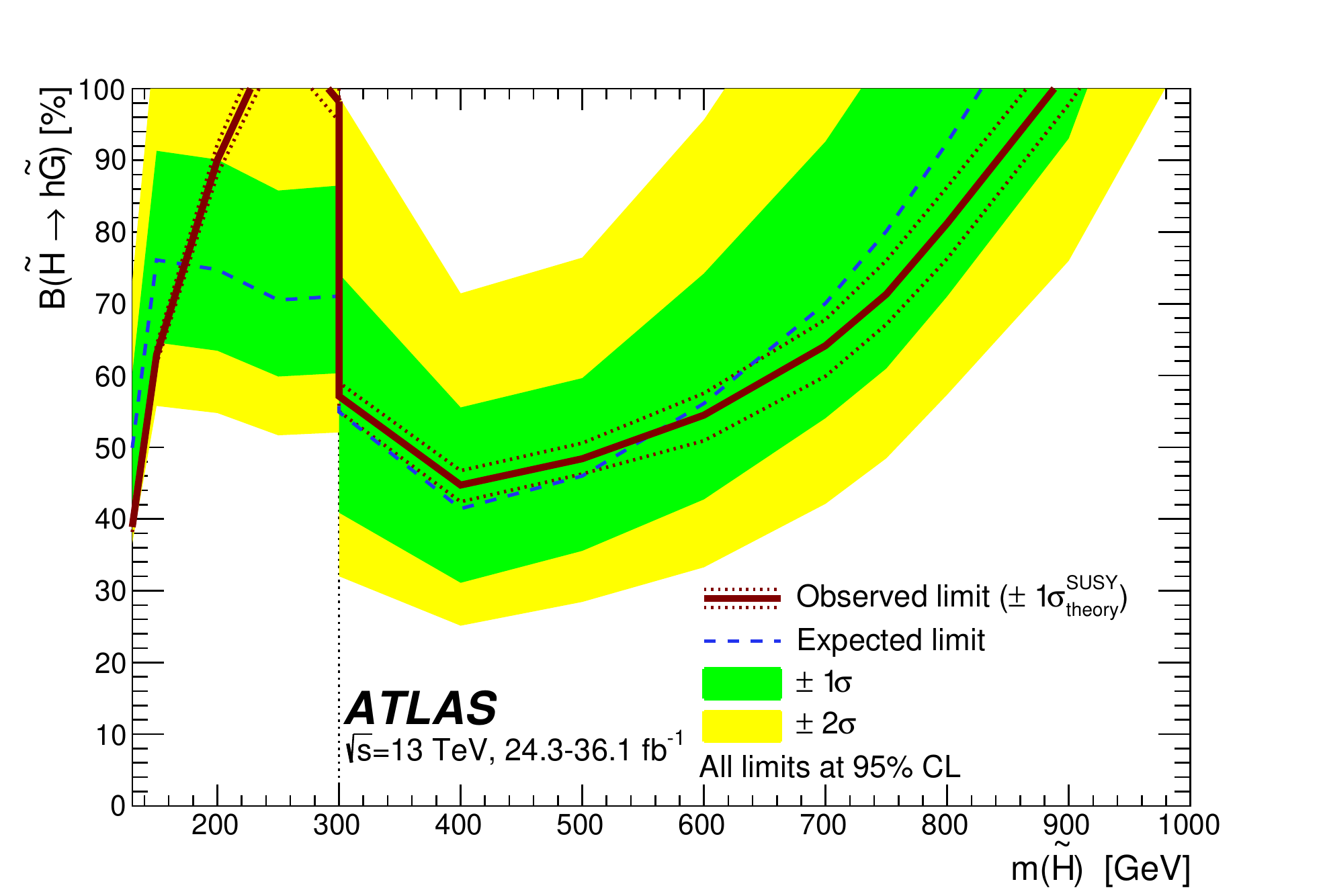}\label{fig:exclusion_br}}
\caption{Exclusion limits on \hino\ pair production. In both interpretations, the results of the low-mass analysis are used below $\mhino = 300$ GeV, while those of the high-mass analysis are used above. In all cases the \gravino is assumed to be nearly massless. The figure shows (a) the observed (solid) vs expected (dashed) 95\% upper limits on the \hino\ pair production cross-section as a function of \mhino.  The 1$\sigma$ and 2$\sigma$ uncertainty bands on the expected limit are shown as green and yellow, respectively. The theory cross-section and its uncertainty are shown in the solid and shaded red curve. The bottom panel shows the ratio of the observed and expected limits with the theory cross-section. The figure also shows (b) the observed (solid) vs expected (dashed) 95\% limits in the \mhino\ vs $B(\hino\rightarrow h \tilde{G})$ plane, where $B(\hino\rightarrow h \tilde{G})$ denotes the branching ratio for the decay $\hino \rightarrow h \gravino$. The 1$\sigma$ uncertainty band is overlaid in green and the 2$\sigma$ in yellow. The regions above the lines are excluded by the analyses.}
\label{fig:exclusion}
\end{figure}

 
\FloatBarrier
 
\section{Conclusions}
\label{sec:conclusion}
A search for pair-produced degenerate higgsinos decaying via Higgs bosons to gravitinos has been performed. LHC $pp$
collision data from the full 2015 and 2016 data-taking periods are studied with an analysis targeting high-mass signals utilizing \met triggers, corresponding to an integrated
luminosity of  36.1~\ifb\ collected at $\rts =13\ \tev$ by the ATLAS detector, 24.3~\ifb\ of which is also used by an analysis utilizing $b$-jet triggers targeting low-mass signals.
Each analysis uses multiple signal regions to maximize sensitivity to the signal models under study. The signal regions require several high-$\pt$ jets,
of which at least three must be $b$-tagged; $\met$; and zero leptons.
For the high-mass analysis, the background is dominated by $\ttbar$+jets production,
which is estimated by MC simulation, after normalizing the event rate
in dedicated control regions.
For the low-mass analysis, the background is dominated by multijet production and is estimated directly from the data.
No excess is found above the predicted background in any of the signal regions. Model-independent limits are
set on the visible cross-section for new physics processes. Exclusion limits are set as a function of the mass of the higgsino; masses between 130 GeV and 230 GeV and between 290 GeV and 880 GeV are excluded at 95\% confidence level.
The results are also interpreted in a model with variable
branching ratios of higgsino decays to a Higgs or $Z$ boson and a gravitino: branching ratios to Higgs boson decays as low as 45\% are excluded for $\mhino \approx 400$ GeV.

\section*{Acknowledgments}
 
 
We thank CERN for the very successful operation of the LHC, as well as the
support staff from our institutions without whom ATLAS could not be
operated efficiently.
 
We acknowledge the support of ANPCyT, Argentina; YerPhI, Armenia; ARC, Australia; BMWFW and FWF, Austria; ANAS, Azerbaijan; SSTC, Belarus; CNPq and FAPESP, Brazil; NSERC, NRC and CFI, Canada; CERN; CONICYT, Chile; CAS, MOST and NSFC, China; COLCIENCIAS, Colombia; MSMT CR, MPO CR and VSC CR, Czech Republic; DNRF and DNSRC, Denmark; IN2P3-CNRS, CEA-DRF/IRFU, France; SRNSFG, Georgia; BMBF, HGF, and MPG, Germany; GSRT, Greece; RGC, Hong Kong SAR, China; ISF, I-CORE and Benoziyo Center, Israel; INFN, Italy; MEXT and JSPS, Japan; CNRST, Morocco; NWO, Netherlands; RCN, Norway; MNiSW and NCN, Poland; FCT, Portugal; MNE/IFA, Romania; MES of Russia and NRC KI, Russian Federation; JINR; MESTD, Serbia; MSSR, Slovakia; ARRS and MIZ\v{S}, Slovenia; DST/NRF, South Africa; MINECO, Spain; SRC and Wallenberg Foundation, Sweden; SERI, SNSF and Cantons of Bern and Geneva, Switzerland; MOST, Taiwan; TAEK, Turkey; STFC, United Kingdom; DOE and NSF, United States of America. In addition, individual groups and members have received support from BCKDF, the Canada Council, CANARIE, CRC, Compute Canada, FQRNT, and the Ontario Innovation Trust, Canada; EPLANET, ERC, ERDF, FP7, Horizon 2020 and Marie Sk{\l}odowska-Curie Actions, European Union; Investissements d'Avenir Labex and Idex, ANR, R{\'e}gion Auvergne and Fondation Partager le Savoir, France; DFG and AvH Foundation, Germany; Herakleitos, Thales and Aristeia programmes co-financed by EU-ESF and the Greek NSRF; BSF, GIF and Minerva, Israel; BRF, Norway; CERCA Programme Generalitat de Catalunya, Generalitat Valenciana, Spain; the Royal Society and Leverhulme Trust, United Kingdom.
 
The crucial computing support from all WLCG partners is acknowledged gratefully, in particular from CERN, the ATLAS Tier-1 facilities at TRIUMF (Canada), NDGF (Denmark, Norway, Sweden), CC-IN2P3 (France), KIT/GridKA (Germany), INFN-CNAF (Italy), NL-T1 (Netherlands), PIC (Spain), ASGC (Taiwan), RAL (UK) and BNL (USA), the Tier-2 facilities worldwide and large non-WLCG resource providers. Major contributors of computing resources are listed in Ref.~\cite{ATL-GEN-PUB-2016-002}.
 

\printbibliography
 
\clearpage
 
\begin{flushleft}
{\Large The ATLAS Collaboration}

\bigskip

M.~Aaboud$^\textrm{\scriptsize 34d}$,    
G.~Aad$^\textrm{\scriptsize 99}$,    
B.~Abbott$^\textrm{\scriptsize 125}$,    
O.~Abdinov$^\textrm{\scriptsize 13,*}$,    
B.~Abeloos$^\textrm{\scriptsize 129}$,    
S.H.~Abidi$^\textrm{\scriptsize 165}$,    
O.S.~AbouZeid$^\textrm{\scriptsize 143}$,    
N.L.~Abraham$^\textrm{\scriptsize 153}$,    
H.~Abramowicz$^\textrm{\scriptsize 159}$,    
H.~Abreu$^\textrm{\scriptsize 158}$
% ...
\bigskip
\\

% ...
$^{13}$Institute of Physics, Azerbaijan Academy of Sciences, Baku; Azerbaijan.\\
% ...
$^{34}$$^{(a)}$Facult\'e des Sciences Ain Chock, R\'eseau Universitaire de Physique des Hautes Energies - Universit\'e Hassan II, Casablanca;$^{(b)}$Centre National de l'Energie des Sciences Techniques Nucleaires (CNESTEN), Rabat;$^{(c)}$Facult\'e des Sciences Semlalia, Universit\'e Cadi Ayyad, LPHEA-Marrakech;$^{(d)}$Facult\'e des Sciences, Universit\'e Mohamed Premier and LPTPM, Oujda;$^{(e)}$Facult\'e des sciences, Universit\'e Mohammed V, Rabat; Morocco.\\
% ...
$^{99}$CPPM, Aix-Marseille Universit\'e, CNRS/IN2P3, Marseille; France.\\
% ...
$^{125}$Homer L. Dodge Department of Physics and Astronomy, University of Oklahoma, Norman OK; United States of America.\\
% ...
$^{129}$LAL, Universit\'e Paris-Sud, CNRS/IN2P3, Universit\'e Paris-Saclay, Orsay; France.\\
% ...
$^{143}$Santa Cruz Institute for Particle Physics, University of California Santa Cruz, Santa Cruz CA; United States of America.\\
% ...
$^{153}$Department of Physics and Astronomy, University of Sussex, Brighton; United Kingdom.\\
% ...
$^{158}$Department of Physics, Technion, Israel Institute of Technology, Haifa; Israel.\\
$^{159}$Raymond and Beverly Sackler School of Physics and Astronomy, Tel Aviv University, Tel Aviv; Israel.\\
% ...
$^{165}$Department of Physics, University of Toronto, Toronto ON; Canada.\\
% ...
$^{a}$ Also at  Department of Physics, University of Malaya, Kuala Lumpur; Malaysia.\\
$^{b}$ Also at Borough of Manhattan Community College, City University of New York, NY; United States of America.\\
$^{c}$ Also at California State University, East Bay; United States of America.\\
$^{d}$ Also at Centre for High Performance Computing, CSIR Campus, Rosebank, Cape Town; South Africa.\\
$^{e}$ Also at CERN, Geneva; Switzerland.\\
% ...
$^{*}$ Deceased

\end{flushleft}


\end{document}